\documentclass{article}
\usepackage{natbib}
\usepackage{graphicx}
\usepackage{palatino}
\usepackage{amsmath}
\usepackage{amsfonts}
\usepackage{amssymb}
\usepackage[T1]{fontenc}
\usepackage{pdfsync}
\usepackage{hyperref}
\DeclareMathOperator{\vect}{vec}
\usepackage{multirow} 
\usepackage{subcaption}

\newtheorem{Remark}{Remark}[section]
\newtheorem{Lemma}{Lemma}[section]

\newtheorem{Theorem}{Theorem}[section]
\newtheorem{Proposition}{Proposition}[section]

\newcommand{\mathbold}[1]{\mbox{\boldmath $\bf#1$}}

\begin{document}

\title{Multivariate Count  Autoregression}
\author{Paul Doukhan$^{1}$~~Konstantinos Fokianos$^{2}$  ~~~~B{\aa}rd St{\o}ve$^{3}$~~~~~~Dag Tj{\o}stheim$^{3}$ \\  \\
$^{1}$Department of Mathematics, University of Cergy-Pontoise\\
e-mail: doukhan@u-cergy.fr \\ \\
$^{2}$Department of Mathematics \& Statistics, University of Cyprus \\
e-mail: fokianos@ucy.ac.cy \\ \\
$^{3}$Department of Mathematics, University of Bergen\\
e-mail:\{{Bard.Stove, Dag.Tjostheim\}}@math.uib.no}
\date{First Version: April 2017}
\maketitle

\begin{abstract}
\noindent
We are studying  the problems of modeling and inference for multivariate count time series data with Poisson marginals.
The focus is  on linear and log-linear models. For studying the properties of such processes
we develop a novel  conceptual framework which is based on copulas.
However, our approach  does not impose the  copula  on a vector of counts; instead the joint distribution
is determined by imposing a copula function  on  a vector of associated  continuous random variables. This specific
construction avoids conceptual difficulties resulting from  the joint distribution of discrete random variables yet it keeps the
properties of the Poisson process marginally.   We employ  Markov chain theory and the
notion of weak dependence to study ergodicity and stationarity of the models we consider. We obtain easily verifiable
conditions for both linear and log-linear models under both theoretical frameworks.
Suitable estimating equations are suggested for  estimating unknown model parameters. The
large sample properties of the resulting estimators are studied in detail.
The work concludes with some simulations and a real data example.
\end{abstract}

\noindent
{\textbf{Keywords:} {autocorrelation, copula, ergodicity,
generalized linear models, perturbation, prediction, stationarity, volatility.}

\tableofcontents

\section{Introduction}

Modeling and inference of multivariate count time series is an important topic for research as such data can
be observed in several applications; see  \citet{Paul2008} for a medical application,
\citet{BoudreaultandCharpentier(2011)} who study the number of earthquake occurrences,
\citet{PedeliandKarlis(2013)} for a financial application and more recently \citet{Ravishankeretal(2015)} for a marketing application.
Some early work in  this research direction can be found in   \citet{FrankeandRao(1995)} and \citet{Latour(1997)}. The interested reader is referred
to the review paper by  \citet{Karlis(2015)},  for further details.

The available literature   shows that there  exist  three main approaches taken
towards the problem of modeling and inference
for multivariate count time series. The first approach is based on the theory of integer autoregressive  (INAR) models and was initiated by
\citet{FrankeandRao(1995)} and \cite{Latour(1997)}. This work was further developed by \citet{PedeliandKarlis(2013b),PedeliandKarlis(2013)}
and a  review of this methodology has been recently given by \citet{Karlis(2015)}. Estimation for these models is based on least
squares methodology and likelihood based methods. However, even in the context of univariate INAR models, likelihood theory is quite
cumbersome, especially for higher order models.
Therefore, this class of  models, which  is adequate to describe some simple data structures, still poses challenges in terms of estimation
(and prediction) especially when the model  order is large.

The second class of models that have been proposed for the analysis of count time series models, is that of parameter driven
models. Recall that a parameter driven model (according to the broad categorization introduced by \citet{Cox(1981)}) is a model whose
dynamics are driven by an unobserved process. In this case,  state space models  for multivariate count time series were studied by
\citet{Jorgensenetall(1996)} and   \citet{Jungetal(2011)} who suggested a factor model for the analysis of  multivariate count time
series; see also \citet{Ravishankeretal(2014), Ravishankeretal(2015)}, among others, for more recent contributions.

The aim of our contribution is to study models that fall within the class of observation driven models; that is models whose
dynamics  evolve according to past values of the process itself plus some noise. This is the case of the usual autoregressive models.
In particular, observation driven models for count time series have been studied by \citet{DavisandDunsmurandWang(2000)},
\citet{Fokianosetal(2009)},  \citet{FokianosandTjostheim(2011)}
\citet{DavisandLiu(2012)}, among
others.
There is a growing recent literature in the topic of modeling and inference for observation driven models for multivariate  count time series; see \citet{HeinenandRegifo(2007)},
\citet{Liu(2012)}, \citet{Andreassen(2013)}, \citet{Ahmad(2016)} and \citet{Leeetal(2017)}, for instance. These studies are mainly concerned with linear model specifications. Although
the linear model is adequate  for studying properties of the multivariate process, it may not always be a natural candidate for count data analysis.
The log-linear model is more appropriate, in our  view, for general modeling of count time series. Some desirable properties of log-linear models
are the ease of including  covariates, incorporation of positive/negative correlation and avoiding parameter boundary problems (see \citet{FokianosandTjostheim(2011)}).
In fact, a  log-linear model corresponds to the canonical link Poisson  regression model for count data analysis (\citet{McCullaghandNelder(1989)}).

\noindent
Besides modeling issues, another   obstacle for the analysis of count time series is the choice of the joint count  distribution.
Indeed, there are numerous proposals
available in the literature generalizing the univariate Poisson probability mass function (pmf); some of these are reviewed in the previous
references. However, the  pmf  of a multivariate Poisson discrete random vector is usually of  quite complicated functional form and therefore
maximum likelihood based inference can be quite challenging (theoretically and numerically).
Generally speaking, the choice of the joint  distribution for multivariate count data is
quite an interesting topic but in this work we have chosen to address this problem by suggesting a copula based construction.
Instead of imposing a copula function on a vector of discrete random variables, we argue, based on Poisson process properties, that
it can be introduced on a vector of continuous random variables. In this way, we avoid some technical difficulties and we propose
a plausible data generating process which keeps intact the properties of the Poisson properties, marginally.
Having resolved the problems of data generating process and given a model, we  suggest suitable estimating functions to
estimate the unknown parameters. The main goals of this work are summarized by the following:
\begin{enumerate}
\item Develop a conceptual framework for studying count time series  with  Poisson process marginally. As it was explained
earlier, one of the problems posed in this setup, is the choice of the joint count  distribution. We resolve this issue  by imposing
a copula structure for accommodating dependence, yet the properties of Poisson processes are kept marginally.
\item Give conditions for ergodicity and stationarity of both linear and log-linear models. The preferred methodologies are those of Markov
chain theory (employing a perturbation approach) and theory of weak dependence.  Although the linear model was treated by \citet{Liu(2012)} in a parametric joint Poisson
framework, we relax these conditions considerably when using the perturbation approach. For the  log-linear model case,  these conditions are new.
\item Furthermore, we suggest a class of estimating functions for inference. As it was discussed earlier, the specification of a joint
distribution of a count vector poses several challenges. We overcome this obstacle by suggesting appropriate  estimating functions which still
deliver consistent and asymptotically normally distributed estimators.
\end{enumerate}

As a final remark we discuss the challenges associated with  the problem of showing   stationarity and ergodicity
of count time series.
The main obstacle--see \citet{Neumann(2010)} and \citet{Tjostheim(2012), Tjostheim(2015)}--is that the process itself
consists of integer valued random variables; however the mean process takes values on the positive real line which creates
difficulties in  proving  ergodicity of the observed process (see \citet{Andrews(1984)} for a related situation).
The study of theoretical properties of these models was initiated by the perturbation method suggested in \citet{Fokianosetal(2009)} and
was further developed in \cite{Neumann(2010)} ($\beta$-mixing), \citep{Doukhanetal(2011)} (weak dependence approach, see \citet{DoukhanandLouhichi(1999)}),
\citet{Doucetal.(2012)} (Markov chain theory without irreducibility assumptions) and \citet{Wangetal(2014)} (based on the theory of $e$-chains; see \citet{MeynandTweedie(1993)}).
We note that \cite{Doukhanetal2012b} study the relation between weak dependence coefficients defined in \citet{DoukhanandLouhichi(1999)}) and
the strong mixing coefficients (\citet{Rosenblatt(1956)}) for the case of integer valued count time series. As it was mentioned before, we will be employing Markov
chain theory (using the  perturbation approach) and the notion of  weak dependence  for studying both linear and log-linear models. The end  results obtained by either
approach are identical when there is no feedback process in the model; however these results change when a hidden process is included in the model.

The paper is organized as follows: Section \ref{sec:models} discusses the basic modeling approach that we take towards
modeling multivariate count time series. The copula structure which is imposed introduces dependence but without affecting
the properties of the marginal Poisson processes. We will consider both a linear and a log-linear model. Section \ref{sec:ergodicity}
gives the results about ergodic and stationary properties of the linear and log-linear models.  Section \ref{Sec:QMLE} discusses Quasi Maximum Likelihood
inference (QMLE) and shows that the resulting estimators are  consistent and asymptotically normal.
Section \ref{sec:examples} presents a limited simulation study and a real data examples.
The paper concludes with a discussion and an appendix which contains the proofs of main results and supplementary material.

\section{Model Assumptions}
\label{sec:models}

In what follows we  assume that $\{{\bf Y}_{t} = (Y_{i, t}), ~i=1,\ldots,p,  ~t=1,2\ldots, \}$ denotes  a $p$--dimensional   count time series.
Let   $\{ {\mathbold \lambda }_{t} = (\lambda_{i, t}), ~i=1,\ldots,p, t=1,2,\ldots  \}$ be the
corresponding $p$-dimensional intensity process   and
${\cal F}_{t}^{\mathbold{Y}, \mathbold{\lambda}}$ the $\sigma$--field generated
by $\{\mathbold{Y}_{0}, \cdots, \mathbold{Y}_{t},  \mathbold{\lambda}_{0} \}$ with  ${\mathbold \lambda}_{0}$ being a $p$-dimensional
vector denoting  the starting value of  $\{ \mathbold{\lambda}_{t} \}$. With this notation, the intensity process is given by
$ {\mathbold \lambda}_{t} =\mbox{E}[\mathbold{Y}_{t} \mid {\cal F}_{t}^{\mathbold{Y}, \mathbold{\lambda}}]$.
We  will be studying  two autoregressive models   for
multivariate count time series analysis;  the linear and log-linear models which are direct extensions of their univariate
counterparts.

\noindent
The linear model is defined by assuming that for each  $i=1,2,\ldots,p$,
\begin{eqnarray}
Y_{i,t} \mid {\cal F}_{t-1}^{\mathbold{Y}, \mathbold{\lambda}}  ~~ \mbox{is marginally} ~~ \mbox{Poisson}(\lambda_{i, t}),  ~~
{\mathbold \lambda}_{t}  =   {\bf d} + {\bf A} {\mathbold \lambda}_{t-1} + {\bf B} {\bf Y}_{t-1},
\label{linear model mult}
\end{eqnarray}
where ${\bf d}$ is a $p$-dimensional vector and $ {\bf A}$, ${\bf B}$ are $p \times p$ unknown  matrices. \emph{The elements of
${\bf d}$,  $ {\bf A}$ and  $ {\bf B}$ are assumed to be positive for ensuring positivity of $\lambda_{i, t}$.}
Model \eqref{linear model mult} generalizes naturally the  linear autoregressive model  discussed by
\citet{RydbergandShephard(2000)}, \citet{Heinen(2003)}, \citet{Ferlandetal(2006)} and \citet{Fokianosetal(2009)}, among others.
The log-linear model that we consider is  the  multivariate analogue
of the univariate  log-linear model proposed by \citet{FokianosandTjostheim(2011)} (see also \citet{Woodardetall(2010)} and \citet{Doucetal.(2012)});
more precisely assume that for each  $i=1,2,\ldots,p$,
\begin{eqnarray}
Y_{i,t} \mid {\cal F}_{t}^{\mathbold{Y}, \mathbold{\lambda}} ~~ \mbox{is marginally} ~~ \mbox{Poisson}(\lambda_{i, t}),~~
{\mathbold \nu}_{t}  =   {\bf d} + {\bf A} {\mathbold \nu}_{t-1} + {\bf B} \log({\bf Y}_{t-1}+{\bf 1}_{p}),
\label{log-linear model mult}
\end{eqnarray}
where ${\mathbold \nu}_{t} \equiv \log {\mathbold \lambda}_{t}$
is defined componentwise (i.e. $\nu_{i,t}= \log \lambda_{i,t}$) and ${\bf 1}_{p}$ denotes the $p$--dimensional
vector which consists of ones. \emph{In the case of \eqref{log-linear model mult}, we do not impose any positivity constraints on
the parameters ${\bf d}$,  $ {\bf A}$ and  $ {\bf B}$; this is an important argument favoring the log-linear model.}
We will examine aspects  of  both models.  The log-linear model \eqref{log-linear model mult}
is expected to be a better candidate for count data  observed jointly with some other covariate time series or where negative correlation is observed.

\noindent
A fundamental problem in the analysis of multivariate count data is the specification of joint distribution of the counts. There
are numerous proposals made in the literature aiming on generalizing the univariate  Poisson assumption to the multivariate case
but the resulting joint distributions
are quite complex for likelihood based inference. For instance, a possible construction can be based on independent Poisson
random variables  or on copulas  and mixture models (see  \citet[Ch. 37]{Johnsonetal(1997)}, \citet[Sec 7.2]{Joe(1997)}). However, the resulting
functional form of the joint pmf  is  complicated and therefore the log-likelihood function
cannot be calculated analytically  (or, sometimes, even approximated).

We propose a quite different approach. Consider the
first equation of \eqref{linear model mult} (but the same  discussion applies to \eqref{log-linear model mult} subject to minor modifications).
It  implies  that each component $Y_{i,t}$ is \emph{marginally} a Poisson process.
But the joint distribution of the vector $\{  {\bf Y}_{t} \}$ is not necessarily distributed as a
multivariate Poisson   random variable. Our general construction, as outlined below, allows
for arbitrary dependence among the marginal Poisson components by utilizing fundamental  properties of the Poisson process.
We give a detailed account of the data generating process.
Suppose that $\mathbold{\lambda}_{0}=(\lambda_{1,0}, \ldots, \lambda_{p,0})$  is some starting value. Then consider
the following data generating mechanism:
\begin{enumerate}
\item Let ${\mathbold U}_{l}=(U_{1 ,l}, \ldots, U_{p,l})$ for $l=1,2,\ldots, K$, be a sample from a $p$-dimensional
copula $C(u_{1}, \ldots, u_{p})$. Then $U_{i,l}$, $l=1,2,\ldots, K$ follow marginally the uniform distribution on $(0,1)$, for
 $i=1,2,\ldots,p$.
\item Consider  the transformation
\begin{eqnarray*}
X_{i, l} = -\frac{ \log U_{i, l}}{\lambda_{i,0}}, ~~~i=1,2, \ldots, p.
\end{eqnarray*}
Then, the marginal distribution of $X_{i,l}$, $l=1,2,\ldots, K$
is exponential with parameter $\lambda_{i,0}$, $i=1,2,\ldots,p$.
\item
Define now (taking $K$ large enough)
\begin{eqnarray*}
Y_{i,0} = \max_{1 \leq k \leq K} \left\{ \sum_{l=1}^{k} X_{i, l} \leq  1 \right\}, ~~~i=1,2,\ldots,p.
\end{eqnarray*}
Then ${\bf Y}_{0}=(Y_{1,0}, \ldots, Y_{p,0})$ is marginally  a set of first values of a  Poisson process with parameter $\mathbold{\lambda}_{0}$.
\item Use model \eqref{linear model mult}  (respectively \eqref{log-linear model mult}) to obtain $\mathbold{\lambda}_{1}$.
\item Return back to step 1 to obtain ${\bf Y}_{1}$, and so on.
\end{enumerate}

\noindent
The aforementioned   construction of the joint distribution of the counts
imposes the dependence among the components of the vector process $\{{\bf Y}_{t}\}$ by taking advantage of a
\emph{copula structure on the waiting times of the Poisson process}. This can be extended to other marginal processes if they can be generated by continuous inter arrival times.
Equivalently, the copula is imposed on the uniform random
variables generating the exponential waiting times. Such an approach  does not pose any  problems on obtaining the
joint distribution of the random vector $\{{\bf Y}_{t}\}$ which is composed of discrete valued random variables.
The copula is defined
uniquely for continuous multivariate random variables (compare with \citet{HeinenandRegifo(2007)} and for a lucid discussion about copula for discrete multivariate
 distributions, see \citet{GenestandNeslehova(2007)}). Hence, the first equation of
model \eqref{linear model mult} can be restated as
\begin{eqnarray}
{\mathbold Y}_{t} ={\bf{N}}_{t}( {\mathbold{\lambda}}_{t}), ~~~
{\mathbold{\lambda}}_{t}  =   {\bf d} + {\bf A} {\mathbold{\lambda}}_{t-1} + {\bf B} {\bf{Y}}_{t-1}
\label{linear model in terms of Poisson}
\end{eqnarray}
where $\{ {\bf N}_{t} \}$ is a sequence of independent $p$-variate copula--Poisson processes which counts the number of events
in $[0, \lambda_{1,t}] \times \ldots \times [0, \lambda_{p,t}]$.  Along the lines of introducing   \eqref{linear model in terms of Poisson} we
also  define the  multivariate log--linear model \eqref{log-linear model mult} by
\begin{eqnarray}
{\mathbold Y}_{t} ={\bf{N}}_{t}( {\mathbold{\nu}}_{t}), ~~~
{\mathbold{\nu}}_{t}  =   {\bf d} + {\bf A} {\mathbold{\nu}}_{t-1} + {\bf B} \log({\bf{Y}}_{t-1}+{\bf 1}_{p})
\label{loglinear model in terms of Poisson}
\end{eqnarray}
recalling that  ${\mathbold{\nu}}_{t} = \log {\mathbold{\lambda}}_{t}$  is defined componentwise and ${\bf 1}_{p}$ denotes the $p$--dimensional
vector which consists of ones. The process $\{ {\bf N}_{t} \}$ denotes as before
a sequence of independent $p$-variate copula--Poisson processes which counts the number of events
in $[0, \exp(\nu_{1,t})] \times \ldots \times [0, \exp(\nu_{p,t})]$.

\noindent
It is instructive to consider model \eqref{linear model in terms of Poisson} in more detail because its structure is closely  related
to the theory of GARCH models, \citet{Bollerslev(1986)}. Observe that each component of the vector-process
$\{\mathbold{Y}_{t}\}$ is distributed as a Poisson random variable. But
the mean of a Poisson random variable equals its variance; therefore  model \eqref{linear model in terms of Poisson} resembles some structure of multivariate
GARCH model, see \citet{Lutkepohl(2005)} and \citet{FrancqandZakoian(2010)}.
Consider   $p=2$, for example. Then the second equation of \eqref{linear model in terms of Poisson} becomes
\begin{eqnarray*}
\lambda_{1,t} & = &  d_{1} + a_{11} \lambda_{1,t-1} + a_{12} \lambda_{2, t-1}+ b_{11} Y_{1, t-1}+ b_{12} Y_{2, t-1}, \\
\lambda_{2,t} & = &  d_{2} + a_{21} \lambda_{1,t-1} + a_{22} \lambda_{2, t-1}+ b_{21} Y_{1, t-1}+ b_{22} Y_{2, t-1},
\end{eqnarray*}
where $d_{i}$ is the $i$th element of $\mathbold{d}$ and $a_{ij}$ ($b_{ij}$, respectively) is the $(i,j)$th element of
$\mathbold{A}$ (\mathbold{B}, respectively). We can give the following interpretation to model parameters. When $a_{12}=b_{12}=0$,
then $\lambda_{1t}$ depends only on its own past. If this is not true, then the parameters denote the linear dependence of $\lambda_{1t}$
on $\lambda_{2, t-1}$ and $Y_{2, t-1}$ in the presence of  $\lambda_{1, t-1}$ and $Y_{1, t-1}$. Similar results hold when  $a_{21}=b_{21}=0$
and the previous discussion applies to the case of \eqref{loglinear model in terms of Poisson}.

\noindent
This section introduced the  approach we take  towards modeling multivariate count time series.
The next section  discusses the properties of the models we consider. We show that their probabilistic properties
can be studied in the framework of Markov chains and weak dependence.

\section{Ergodicity and Stationarity}
\label{sec:ergodicity}

\noindent
Towards the analysis of models \eqref{linear model in terms of Poisson} and \eqref{loglinear model in terms of Poisson}, we employ the perturbation techniques
as developed by \citet{Fokianosetal(2009)} and \citet{FokianosandTjostheim(2011)}.
In addition, we include  a study which is based on  the notion  of weak dependence (for more, see \citet{DoukhanandLouhichi(1999)} and \citet{Dedeckeretal(2007)}). Both approaches
are employed and compared for obtaining ergodicity and stationarity of \eqref{linear model in terms of Poisson} and \eqref{loglinear model in terms of Poisson}.
In fact, the main goal is to obtain stationarity and ergodicity of the joint process $( \mathbold{Y}_{t}, \mathbold{\lambda}_{t})$.
Such a result is of importance on studying the asymptotic distribution of the quasi-maximum likelihood estimator discussed in Section \ref{Sec:QMLE}. The problem of proving such results for  the joint process is the discreteness
of the component $\mathbold{Y}_{t}$. The perturbation method and the weak dependence approach allows us to bypass successfully this problem and derive
sufficient  conditions for proving
the desired properties of the joint process. Alternative methods to approach this problem have been studied by  \citet{Neumann(2010)},  \citet{Woodardetall(2010)}, \citet{Doucetal.(2012)} and
\citet{Wangetal(2014)}. The recent review articles by  \citet{Tjostheim(2012),Tjostheim(2015)} discuss in detail these issues.
For the specific examples of processes given by \eqref{linear model in terms of Poisson} and \eqref{loglinear model in terms of Poisson} the
sufficient conditions  obtained by the perturbation and weak dependence approach are different; however all proofs are based on
a contraction property  of the process $\{ \mathbold{\lambda}_{t} \}$  (in the case of \eqref{linear model in terms of Poisson}) and
$\{ \mathbold{\nu}_{t} \}$  (in the case of \eqref{loglinear model in terms of Poisson}).  We initialize  the  discussion
with the linear model. We  denote  by $\|{\bf x}\|_{d}= (\sum_{i=1}^{p} |x_{i}|^{d})^{1/d}$ the $l^d$- norm of
a $p$-dimensional vector ${\bf x}$. For an $m \times n$ matrix $\mathbold{A}=(a_{ij})$, ${i=1,\ldots,m, j=1,\ldots,n}$,  we let
$\|| \mathbold{A} \||_{d}$  denote the generalized matrix norm  induced by $\|\cdot\|_{d}$, for  $d \geq 1$. In other words
$\|| \mathbold{A} \||_{d}= \max_{\|{\bf x}\|_{d}=1} \| \mathbold{A} \mathbold{x} \|_{d}$. If $d=1$, then
$\|| \mathbold{A} \||_{1}=\max_{1 \leq j \leq n} \sum_{i=1}^{m} |a_{ij}|$, and when $d=2$,
$\|| \mathbold{A} \||_{2}=\rho^{1/2}( \mathbold{A}^{T} \mathbold{A})$  where $\rho(.)$ denotes the spectral radius
of a matrix. Moreover, the Frobenious norm is denoted by $\||\mathbold{A}\||_{F}= \left(\sum_{i,j} |a_{ij}|^{2} \right)^{1/2}$. If $m=n$, then these norms are matrix norms.

\subsection{Linear Model}

Following \citet{Fokianosetal(2009)}, we introduce the perturbed model
\begin{eqnarray}
{\bf Y}_{t}^{m} ={\bf N}_{t}( {\bf \lambda}_{t}^{m}), ~~
{\mathbold \lambda}_{t}^{m}  =   {\bf d} + {\bf A} {\mathbold \lambda}_{t-1}^{m} + {\bf B} {\bf Y}_{t-1}^{m}+ \mathbold{\epsilon}_{t}^{m},
\label{perturbed linear}
\end{eqnarray}
where $\mathbold{\epsilon}_{t}^{m}=c_{m} {\bf V}_{t}$. Here  the sequence $c_{m}$ is strictly positive  and tends to zero, as
$m \rightarrow \infty$, and ${\bf V}_{t}$ is a $p$-dimensional vector which consists of independent positive random variables
each of which having a bounded support of the form $[0, M]$, for some $M >0$. The introduction of the perturbed process allows
to prove ergodicity and stationarity of the joint process $\{ (\mathbold{Y}^{m}_{t}, \mathbold{\lambda}^{m}_{t}, \mathbold{\epsilon}_{t}^{m}) \}$.
The first result is given by the following proposition:

\begin{Proposition} \rm
Consider model \eqref{perturbed linear} and suppose that $\|| {\bf A} + {\bf B} \||_{2} < 1$. Then the process $\{ \mathbold{\lambda}_{t}^{m}, t >0\}$
is a geometrically ergodic Markov chain with finite $r$'th moments, for any $r >0$. Moreover,
the process $\{ ({\bf Y}_{t}^{m}, {\mathbold \lambda}_{t}^{m}, {\mathbold \epsilon}_{t}), t> 0\}$ is $V_{{\bf Y}, {\mathbold \lambda}, {\mathbold \epsilon}}$
geometrically ergodic Markov chain with $V_{{\bf Y}, {\mathbold \lambda}, {\bf \epsilon}}=1+\|{\bf Y}\|_{2}^{r}+\|{\mathbold \lambda}\|_{2}^{r}+\|{\mathbold\epsilon}\|_{2}^{r}$, $r>0$.
\label{ergodicity linear perturbed}
\end{Proposition}

\noindent
The following results shows that as $c_{m} \rightarrow 0$ as $m \rightarrow \infty$, then the difference between \eqref{linear model in terms of Poisson}
and \eqref{perturbed linear} can be made arbitrary small.

\begin{Lemma} \rm
Consider models \eqref{linear model in terms of Poisson}
and \eqref{perturbed linear}. If $\|| {\bf A}+  {\bf B} \||_{2} < 1$, then the following hold true:
\begin{enumerate}
\item $\|\mbox{E}( {\mathbold \lambda}_{t}^{m}-{\mathbold \lambda}_{t})\|_{2}=\|\mbox{E}( {\bf Y}_{t}^{m}-{\bf Y}_{t})\|_{2} \leq \delta_{1,m}.$
\item $\mbox{E}\| ({\mathbold \lambda}_{t}^{m}-{\mathbold \lambda}_{t})\|_{2}^{2} \leq \delta_{2,m}.$
\item $\mbox{E}\| ( {\bf Y}_{t}^{m}-{\bf Y}_{t})\|_{2}^{2} \leq \delta_{3,m}.$
\end{enumerate}
In the above $\delta_{i,m} \rightarrow 0$, as $m \rightarrow \infty$. In addition, for sufficiently large $m$
\begin{eqnarray*}
\|{\mathbold \lambda}_{t}^{m}-{\mathbold \lambda}_{t}\|_{2} \leq \delta ~~~\mbox{and}~~~\|{\bf Y}_{t}^{m}-{\bf Y}_{t}\|_{2} \leq \delta,
\end{eqnarray*}
almost surely, for any $\delta > 0$.
\label{lemma approx linear}
\end{Lemma}

The above results show that  the condition  $\|| {\bf A} +  {\bf B} \||_{2} < 1$ is sufficient   to guarantee the required
contraction (c.f. Lemma \eqref{lemma approx linear}) and existence of all moments of the joint process
$\{({\bf Y}_{t}, {\mathbold{\lambda}}_{t}) \}$, (see Proposition \eqref{ergodicity linear perturbed}).
In the simple case of a vector autoregressive model with ${\bf A}= \mathbold{0}$ in \eqref{linear model in terms of Poisson},
the condition $\||{\bf B} \||_{2} < 1$ guarantees stationarity and
ergodicity of the process $\{ \bf{Y}_{t} \}$.
This fact is proved by iterating the recursions of the autoregressive model yielding powers of ${\bf B}$.
However, this technique cannot be applied to the general  multivariate  case but it is deduced by Proposition \ref{ergodicity linear perturbed}.
We conjecture that for the general linear multivariate model of the form
\begin{eqnarray*}
{\mathbold \lambda}_{t} & = &  {\bf d} + \sum_{i=1}^{l} {\bf A}_{i} {\mathbold \lambda}_{t-i} + \sum_{j=1}^{q}{\bf B}_{j} {\bf Y}_{t-j},
\end{eqnarray*}
the condition $\sum_{i=1}^{\max(l, q)}\||{\bf A}_{i} +{\bf B}_{i} \||_{2}  < 1$
is sufficient for proving Proposition \ref{ergodicity linear perturbed}.

We turn now to an alternative method; namely we will use the concept of weak dependence  to study the properties of the linear model
\eqref{linear model in terms of Poisson}. This approach
does not require a perturbation argument but the sufficient  conditions obtained are weaker. The proof of this result parallels the proof
of \citet{Doukhanetal(2011)}; we outline some aspects of it in the appendix.

\begin{Proposition} \rm
Consider model \eqref{linear model in terms of Poisson} and suppose that $\|| {\bf A} \||_{1} + \|| {\bf B} \||_{1} < 1$.
Then there exists a unique causal solution $\{(\mathbold{Y}_{t},\mathbold{\lambda}_{t})\}$ to model \eqref{linear model in terms of Poisson} which is stationary, ergodic and satisfies $\mbox{E}\|\mathbold{Y}_{t}\|_{r}^{r} < \infty $ and $\mbox{E}\|\mathbold{\lambda}_{t}\|_{r}^{r}<  \infty $,
for any $r\in\mathbb{N}$.
\label{ergodicity linear perturbed wd}
\end{Proposition}

The closest result reported in the literature analogous to those obtained by  Propositions \ref{ergodicity linear perturbed} and
\ref{ergodicity linear perturbed wd} can be found in \citet[Prop. 4.2.1]{Liu(2012)} which is, in fact, based on the assumption of  a joint multivariate Poisson distribution for the vector of counts.
The author shows that if  there exists a $p\geq 1$ such that $\||\mathbold{A}\||_{p} + 2^{1-(1/p)}\||\mathbold{B}\||_{p} <1$
then the process $\{\mathbold{\lambda}_{t}\}$ is geometrically moment contracting, see \citet{Wu(2011)} for definition. In the case that $p=2$, then the  condition
of Proposition \ref{ergodicity linear perturbed}  improves this result for the perturbed process $\{ \mathbold{\lambda}_{t}^{m} \}$. When $p=1$ we see that the aforementioned condition is reduced to that
proved in Proposition \ref{ergodicity linear perturbed wd}.

\label{pageremarkinarch}
\noindent
As a closing remark, note that \eqref{linear model in terms of Poisson}  can be   iterated as
\begin{eqnarray}
\mathbold{\lambda}_t&=& \mathbold{d}+\mathbold{A} \mathbold{\lambda}_{t-1}+\mathbold{B} \mathbold{Y}_{t-1}
\nonumber \\
& =& \mathbold{d}+\mathbold{A} \mathbold{d}+\mathbold{A}^2\mathbold{\lambda}_{t-2}+\mathbold{A}\mathbold{B} \mathbold{Y}_{t-2}+\mathbold{B}\mathbold{Y}_{t-1}
\nonumber \\
& =& \mathbold{d}+ \mathbold{A} \mathbold{d}+\mathbold{A}^2 \mathbold{d}+\mathbold{A}^3 \mathbold{\lambda}_{t-3}+\mathbold{A}^2 \mathbold{B} \mathbold{Y}_{t-3}+\mathbold{A} \mathbold{B} \mathbold{Y}_{t-2}
+\mathbold{B} \mathbold{Y}_{t-1}
\nonumber \\
&=&\cdots
\nonumber \\
& =&\sum_{j=0}^{k-1}\mathbold{A}^j \mathbold{d}+\mathbold{A}^k \mathbold{\lambda}_{t-k}+\sum_{j=0}^{k-1} \mathbold{A}^j \mathbold{B} \mathbold{Y}_{t-j-1}
\label{infiniteinarrepresentation}
\end{eqnarray}
Assume that $\||\mathbold{A}\||_{2}<1$. Then an  alternative representation of model \eqref{linear model mult} holds,
from a passage to the limit, as  $k\uparrow\infty$,  from  the above equation:
\begin{equation}\label{inarchinfty}
\mathbold{Y}_t=\mathbold{N}_t(\mathbold{\lambda}_t),\qquad \mathbold{\lambda}_t=(\mathbold{I}_{p}-\mathbold{A})^{-1} \mathbold{d}+\sum_{j=0}^{\infty} \mathbold{A}^j \mathbold{B} \mathbold{Y}_{t-j-1}.
\end{equation}
where ${\bf I}_{p}$ is the identity matrix of order $p$.

\noindent
In this case,  the stationarity condition obtained from
\citet{DoukhanandWintenberger(2008)}, as a multivariate variant of \citet{Doukhanetal(2011)}, is given by
\begin{equation}\label{inarchc}
\sum_{j=0}^{\infty}\|| \mathbold{A}^j \mathbold{B}\||_{2} <1.
\end{equation}
This condition is implied from $\||\mathbold{A}\||_{2}+\||\mathbold{B}\||_{2}<1$.  Indeed,  $\|| \mathbold{A}^j \mathbold{B}\||_{2} \le \|| \mathbold{A}\||_{2}^j\cdot\||\mathbold{B}\||_{2}$  and therefore
$$
\sum_{j=0}^{\infty}\||\mathbold{A}^j \mathbold{B}\||_{2}\le \sum_{j=0}^{\infty}\||\mathbold{A}\||^j\cdot\||\mathbold{B}\||_{2}=\frac{\||\mathbold{B}\||_{2}}{1-\||\mathbold{A}_{2}\||}<1.
$$
In other words,   \eqref{inarchc} improves Proposition \ref{ergodicity linear perturbed wd}.
However, if  $\mathbold{A}\mathbold{B}=\mathbold{B}\mathbold{A}$ and if they are  non-negative  definite, then we obtain that $\||\mathbold{A}+\mathbold{B}\||_2=\||\mathbold{A}\||_2+\||\mathbold{B}\||_2$
and then all obtained conditions coincide. To see that holds true, note that when $\mathbold{A}\mathbold{B}=\mathbold{B}\mathbold{A}$ then $\mathbold{A}, \mathbold{B}$
can  be simultaneously reduced in triangular blocks  with the same eigenvalue on each block.


\subsection{Log-linear Model}

We  turn to the study of  the log--linear model \eqref{loglinear model in terms of Poisson}. We introduce again its
perturbed version by
\begin{eqnarray}
{\bf Y}_{t}^{m}={\bf N}_{t}( {\mathbold{\lambda}}_{t}^{m}), ~~
{\mathbold \nu}_{t}^{m}  =  {\bf d} + {\bf A} {\mathbold \nu}_{t-1}^{m} + {\bf B}\log( {\bf Y}_{t-1}^{m}+{\bf 1}_{p})+ \mathbold{\epsilon}_{t}^{m},
\label{perturbed loglinear}
\end{eqnarray}
where the perturbation has the same structure as in  \eqref{perturbed linear}; .
Then, \citet[Lemma A.2]{FokianosandTjostheim(2011)}
show that $\mbox{E}[ (\log(Y_{j, t-1}^{m}+1))^{r} | \nu_{j; t-1}=\nu_{j}] \sim  \nu_{j}^{r}$, $j=1,2,\ldots, p$ and $r> 0$.
Therefore, we can employ  similar arguments as those employed in \citet{FokianosandTjostheim(2011)}
to prove the following results.

\begin{Proposition} \rm
Consider \eqref{perturbed loglinear} and suppose that $\|| {\bf A}\||_{2} + \||{\bf B} \||_{2} < 1$. Then the process $\{ \mathbold{\nu}_{t}^{m}, t >0\}$
is geometrically ergodic Markov chain with finite $r$'th moments, for any $r >0$. Moreover,
the process $\{ ({\bf Y}_{t}^{m}, {\mathbold \nu}_{t}^{m}, {\mathbold \epsilon}_{t}), t> 0\}$ is $V_{{\bf Y}, {\mathbold \nu}, {\mathbold \epsilon}}$
geometrically ergodic Markov chain with $V_{{\bf Y}, {\mathbold \lambda}, {\mathbold  \epsilon}}=1+\|\log ({\bf Y}+\mathbold{1}_{p})\|_{2}^{2r}+\|{\mathbold \nu}\|_{2}^{2r}+
\|{\mathbold \epsilon}\|_{2}^{2r}$, $r>0$.
\label{ergodicity loglinear perturbed}
\end{Proposition}

\noindent
The proof of the above result is omitted. However, we give in the appendix some details about the following approximation lemma.

\begin{Lemma} \rm
Consider models \eqref{loglinear model in terms of Poisson}
and \eqref{perturbed loglinear}. If $\|| {\bf A} \||_{2}+ \|| {\bf B} \||_{2} < 1$, then the following hold true:
\begin{enumerate}
\item $\|\mbox{E}( {\mathbold \nu}_{t}^{m}-{\mathbold \nu}_{t}\|_{2} \rightarrow 0$, as $m \rightarrow \infty$ and
$\|\mbox{E}( {\bf Y}_{t}^{m}-{\bf Y}_{t})\|_{2} \leq \delta_{1,m}.$
\item $\mbox{E}\| ({\mathbold \nu}_{t}^{m} -{\mathbold \nu}_{t})\|_{2}^{2} \leq \delta_{2,m}.$
\item $\mbox{E}\|( {\bf Y}_{t}^{m}-{\bf Y}_{t})\|_{2}^{2} \leq \delta_{3,m}.$
\item  $\mbox{E}\| ({\mathbold \lambda}_{t}^{m} -{\mathbold \lambda}_{t})\|_{2}^{2} \leq \delta_{4,m}.$
\end{enumerate}
In the above $\delta_{i,m} \rightarrow 0$, as $m \rightarrow \infty$. In addition, for sufficiently large $m$
\begin{eqnarray*}
\|{\mathbold \nu}_{t}^{m}-{\mathbold \nu}_{t}\|_{2} \leq \delta ~~~\mbox{and}~~~\|{\bf Y}_{t}^{m}-{\bf Y}_{t}\|_{2} \leq \delta,
\end{eqnarray*}
almost surely, for any $\delta > 0$.
\label{lemma approx loglinear}
\end{Lemma}

\noindent
We see that the condition  $\|| {\bf A} +  {\bf B} \||_{2} < 1$  obtained for the  linear model  \eqref{linear model in terms of Poisson}
is not   implied  by the condition $\|| {\bf A}\||_{2}+ \||{\bf B} \||_{2} <1 $
which was found for the log-linear model. Recall that in the case of the linear model \eqref{linear model in terms of Poisson} all parameters are assumed to be
positive for ensuring that the components of $\mathbold{\lambda}_{t}$ are positive. This is not necessary for the log-linear model case.
Closing this section, we note that the weak dependence approach delivers
a similar condition.

\begin{Proposition} \rm
Consider model \eqref{loglinear model in terms of Poisson} and suppose that $\|| {\bf A} \||_{1} + \|| {\bf B} \||_{1} < 1$.
Then there exists a unique causal solution $\{(\mathbold{Y}_{t},\mathbold{\nu}_{t})\}$ to model \eqref{log-linear model mult} which is stationary, ergodic and satisfies $\mbox{E}\|\log(\mathbold{Y}_{t}+\mathbold{1}_{p})\|_{r}^{r} < \infty $ and
$\mbox{E}\|\mathbold{\nu}_{t}\|_{r}^{r} < \infty$ and $\mbox{E}[ \exp(r \|\mathbold{\nu}_{t}\|_{1})] < \infty$ for any $r \in \mathbb{N}$.
\label{ergodicity loglinear perturbed wd}
\end{Proposition}

The same remarks made for the linear model \eqref{linear model in terms of Poisson} in page \pageref{pageremarkinarch} hold true for the case of the log-linear model \eqref{loglinear model in terms of Poisson}.
Indeed,  note that the infinite representation is still valid by replacing $\mathbold{\lambda}_{t}$ by
$\mathbold{\nu}_{t}$ and $\mathbold{Y}_{t}$ by $\log( \mathbold{Y}_{t}+ \mathbold{1}_{p})$. Hence, \eqref{inarchc} asserts stationarity and weak dependence for the log-linear model.
In both cases we were not able to prove the conjecture that  $\|| {\bf A} +  {\bf B} \||_{2} < 1$ implies  weak dependence. However, \eqref{inarchc}  improves  on  the results of Lemmas
\ref{ergodicity linear perturbed wd} and \ref{ergodicity loglinear perturbed wd}.

\section{Quasi-Likelihood Inference}
\label{Sec:QMLE}

Suppose that $\{ {\bf Y}_{t}, t=1,2,\ldots,n \}$  is an available sample  from a  count time series and let the vector of unknown parameters
to be denoted by   ${\mathbold \theta }$; that is ${\mathbold \theta }=({\bf d}^{T}, \vect^{T}({\bf A}), \vect^{T}({\bf B}))$, where $\vect(\cdot)$
denote the $\vect$ operator and  $\dim({\bf \theta }) \equiv d = p(1+2p)$.
The general approach that we take towards the estimation problem is based on the theory of estimating functions as outlined by
\citet{LiangandZeger(1986)} for longitudinal data analysis
and \citet{BasawaandRao(1980)}, \citet{Heyde(1997)}, among others,   for stochastic processes. We will be considering
the following conditional quasi--likelihood function  for the parameter vector  ${\mathbold\theta }$,
\begin{eqnarray*}
L({\mathbold \theta}) = \prod_{t=1}^{n} \prod_{i=1}^{p}\Bigl\{ \frac{ \exp(-\lambda_{i,t}({\mathbold \theta}))\lambda_{i,t}^{y_{i,t}}({\mathbold \theta})}{y_{i,t}!} \Bigr\}.
\end{eqnarray*}
This is  equivalent to  considering  model \eqref{linear model mult} (and \eqref{log-linear model mult})
under the assumption of contemporaneous independence among time series.  This assumption simplifies considerably computation of estimators and their standard
errors. At the same time, our approach is based on some simple assumptions which guarantee consistency and asymptotic normality of the
resulting estimator (see \citet{ChristouandFokianos(2014)} and \citet{AhmadandFrancq(2015)} for related recent contributions in the context of
count time series). In fact, the main idea is the correct mean model specification. In other words, if we assume that for a given count time
series and regardless of the true data generating process, there exists a "true" vector of parameters, say ${\mathbold \theta }_{0}$, such that
\eqref{linear model mult} holds (respectively \eqref{log-linear model mult}), then we obtain consistent and asymptotically normally distributed estimators
by maximizing the quasi log-likelihood function \eqref{quasi-loglikelihood}. We point out that \citet{Ahmad(2016)}, independent of us, considered the same approach but his
work neither gives conditions for ergodicity for the models we examine nor does  it consider log-linear multivariate models.
In the following, we give some  details for the linear model case  but  inference  can be
easily developed  for the log--linear model \eqref{log-linear model mult} following  the
same arguments; we will only highlight some different aspects of each model.

\noindent
The quasi  log-likelihood function is equal to
\begin{eqnarray}
l({\mathbold \theta}) = \sum_{t=1}^{n} \sum_{i=1}^{p} \Bigl( y_{i,t} \log \lambda_{i,t}({\mathbold \theta})-\lambda_{i,t}({\mathbold \theta}) \Bigr).
\label{quasi-loglikelihood}
\end{eqnarray}
We denote by  $\widehat{\mathbold{\theta}}\equiv \arg \max_{\mathbold{\theta}} l({\mathbold \theta}), $ the QMLE of $\mathbold{\theta}$.
The score function is given by
\begin{eqnarray}
S_{n}(\mathbold \theta)= \sum_{t=1}^{n} \sum_{i=1}^{p} \Bigl( \frac{y_{i,t}}{\lambda_{i,t}(\mathbold{\theta})}-1 \Bigr)
\frac{\partial \lambda_{i,t}(\mathbold \theta)}{\partial \mathbold{\theta}}
=  \displaystyle \sum_{t=1}^{n} \frac{ \partial {\mathbold{\lambda}}^{T}_{t}(\mathbold{\theta})}{ \partial {\mathbold{\theta}}} {\bf D}^{-1}_{t}
({\mathbold \theta})\Bigl({\bf Y}_{t}- {\mathbold{\lambda}}_{t}({\mathbold{\theta}})\Bigr) \equiv \sum_{t=1}^{n} s_{t}(\mathbold{\theta}),
\label{score independnence}
\end{eqnarray}
where ${\partial {\mathbold{\lambda}}_{t}}/{ \partial {\mathbold{\theta}^{T}}}$ is a $p \times d$ matrix and ${\bf D}_{t}$ is the $p \times p$
diagonal matrix with the $i$'th diagonal element equal to $\lambda_{i,t}({\bf \theta})$, $i=1,2,\ldots,p$. Straightforward differentiation shows
that under model \eqref{linear model mult}, we obtain the following recursions:
\begin{eqnarray}
\frac{ \partial {\mathbold{\lambda}}_{t}}{ \partial {\mathbold{d}^{T}}}  & = & {\bf I}_{p} + {\bf A} \frac{ \partial {\mathbold{\lambda}}_{t-1}}{ \partial {\mathbold{d}}^{T}}, \nonumber\\
\frac{ \partial {\mathbold{\lambda}}_{t}}{ \partial \vect^{T}({\bf A})}    & = & ({\mathbold \lambda}_{t-1} \otimes {\bf I}_{p})^{T} + {\bf A}
\frac{ \partial {\mathbold{\lambda}}_{t-1}}{ \partial \vect^{T}({\bf A}) },  \label{recursions linear} \\
\frac{ \partial {\mathbold{\lambda}}_{t}}{ \partial \vect^{T}({\bf B})}    & = & ({\bf Y}_{t-1} \otimes {\bf I}_{p})^{T} + {\bf A}
\frac{ \partial {\mathbold{\lambda}}_{t-1}}{ \partial \vect^{T}({\bf B}) }, \nonumber
\end{eqnarray}
where  $\otimes$ denotes Kronecker's product.
The   Hessian matrix is given by
\begin{eqnarray}
{\bf H}_{n}(\mathbold{\theta}) & = &
\sum_{t=1}^{n} \sum_{i=1}^{p} \frac{y_{i;t}}{\lambda_{i;t}^{2}(\mathbold{\theta})}
\frac{\partial \lambda_{i,t}(\mathbold{\theta})}{\partial \mathbold{\theta}}
\frac{\partial \lambda_{i,t}(\mathbold{\theta})}{\partial \mathbold{\theta}^{T}}-
\sum_{t=1}^{n} \sum_{i=1}^{p} \Bigl( \frac{y_{i,t}}{\lambda_{i;t}(\mathbold{\theta})}-1 \Bigr)
\frac{\partial^{2} \lambda_{i,t}(\mathbold \theta)}{\partial \mathbold{\theta} \partial \mathbold{\theta}^{T}}.
\label{Hessian}
\end{eqnarray}
Therefore,  the conditional information matrix is equal to
\begin{eqnarray}
{\bf G}_{n}(\mathbold{\theta})& = &
\sum_{t=1}^{n}   \frac{ \partial {\mathbold{\lambda}}^{T}_{t}(\mathbold{\theta})}{ \partial {\mathbold{\theta}}} {\bf D}^{-1}_{t}(\mathbold{\theta})
\mathbold{\Sigma}_{t}(\mathbold{\theta})
{\bf D}^{-1}_{t}(\mathbold{\theta}) \frac{ \partial {\mathbold{\lambda}}_{t}(\mathbold{\theta})}{ \partial {\mathbold{\theta}^{T}}},
\label{information}
\end{eqnarray}
where the matrix $\mathbold{\Sigma}_{t}(\cdot)$ denotes the \emph{true} covariance matrix of the vector $\mathbold{Y}_{t}$. In case
that the process $\{\mathbold{Y}_{t}\}$ consists of uncorrelated components then $ \mathbold{\Sigma}_{t}(\mathbold{\theta})=
{\bf D}_{t}(\mathbold{\theta})$.

We will study the asymptotic properties of the QMLE $\widehat{\mathbold{\theta}}$. By using \citet[Thm 3.2.23]{TaniguchiandKakizawa(2000)} which is based on the work by \citet{KlimkoandNelson(1978)},
we can prove  existence, consistency and asymptotic normality of $\widehat{\mathbold{\theta}}$. Continuous differentiability of the log-likelihood
function, which is guaranteed by the Poisson assumption, is instrumental for  obtaining  these results.
The main problem that we are faced with  is that
we cannot use directly  the sufficient  ergodicity and stationarity   conditions for the unperturbed model to obtain the asymptotic theory (see also \citet{Fokianosetal(2009)}, \citet{FokianosandTjostheim(2011)}
and \citep{Tjostheim(2012),Tjostheim(2015)} for detailed discussion about the issues involved).
Therefore we use  the corresponding conditions  for the perturbed model and then show that the perturbed and unperturbed versions are "close".
Towards this goal  define analogously $S_n^m$ to be the  MQLE score function for the perturbed model
with $(\mathbold{Y}_t,\mathbold{\lambda}_t)$ replaced by $(\mathbold{Y}_t^m,\mathbold{\lambda}_t^m)$. Then, Theorem  \ref{Theorem1} follows immediately after
proving Lemmas \ref{Lemma1qmle}-\ref{Lemma3qmple}  and  taking into account Remark \ref{remark thirdderivativemqle} concerning the third derivative of the log-likelihood function.
Together these results  verify the conditions of \citet[Thm 3.2.23]{TaniguchiandKakizawa(2000)}.

\begin{Lemma}
\normalfont Define the matrices (see \eqref{information limit})
\begin{equation*}
{\mathbold G}^m(\mathbold{\theta})= \mbox{E}\Bigl(s_t^m(\mathbold{\theta}) s_t^m(\mathbold{\theta})^T \Bigr)~~~\text{and}~~~
\mathbold{G}(\mathbold{\theta})= \mbox{E}\Bigl(s_t(\mathbold{\theta}) s_t(\mathbold{\theta})^T \Bigr).
\end{equation*}
Under the assumptions of Theorem  \ref{Theorem1} the above matrices evaluated at the true value $\mathbold{\theta}=\mathbold{\theta}_0$, satisfy
${\mathbold G}^m \rightarrow {\mathbold G}$, as $m \rightarrow \infty$.
\label{Lemma1qmle}
\end{Lemma}

\begin{Lemma}
\normalfont Under the assumptions of Theorem  \ref{Theorem1} the score functions for the perturbed \eqref{perturbed linear} and unperturbed model \eqref{loglinear model in terms of Poisson} evaluated at the true value $\mathbold{\theta}=\mathbold{\theta}_0$ satisfy the following:
\begin{enumerate}
\item $S_n^m/{n} \xrightarrow{\text{a.s}} 0$,
\item $S_n^m/{\sqrt{n}} \xrightarrow{\text{d}} S^m := N(0, {\mathbold G}^m)$,
\item $S^m \xrightarrow{\text{d}} N(0, {\mathbold G})$, as $m \rightarrow \infty$,
\item $\lim_{m\to\infty} \limsup_{n\to\infty} P(||S_n^m-S_n||_{2}>\epsilon \sqrt{n})=0,~~~~\forall \epsilon >0 $.
\end{enumerate}
\label{Lemma2qmle}
\end{Lemma}

\begin{Lemma}
\normalfont
Recall the Hessian matrix defined by \eqref{Hessian},  $\mathbold{H}_{n}$,  and let $ \mathbold{H}^{m}_{n}$
be the Hessian matrix which corresponds to the perturbed model \eqref{perturbed linear}
evaluated at the true value $\mathbold{\theta}=\mathbold{\theta}_0$.
Then, under the assumptions of Theorem \ref{Theorem1}
\begin{enumerate}
\item $\mathbold{H}_n^m \xrightarrow{\text{p}}  \mathbold{H}^m$ as $n \rightarrow \infty$
\item $\lim_{m\to\infty} \limsup_{n\to\infty} P(\||\mathbold{H}_n^m-\mathbold{H}_n\||_{2}>\epsilon n)=0,~~~~\forall \epsilon >0 $.
\end{enumerate}
where $\mathbold{H}$ has been defined by \eqref{Hessian limit} (and analogously for $\mathbold{H}^{m}$). In addition, the matrix $\mathbold{H}$ is positive definite.
\label{Lemma3qmple}
\end{Lemma}

\begin{Theorem}
\normalfont
Consider model \eqref{linear model in terms of Poisson}. Let $\mathbold{\theta} \in \boldsymbol{\Theta} \subset \mathbb{R}^{d}$.
Suppose that $\boldsymbol{\Theta}$ is compact and assume that
the true value $\mathbold{\theta}_{0}$ belongs to the interior of $\boldsymbol{\Theta}$.
Suppose that at the true value $\mathbold{\theta}_0$, the condition of Proposition \ref{ergodicity linear perturbed} hold true.
Then there exists a fixed open neighborhood, say $O(\mathbold{\theta}_0)=\{ \mathbold{\theta}: ~\|\mathbold{\theta}-\mathbold{\theta}_{0}\|_{2} < \delta\}$, of $\mathbold{\theta}_0$
such that with probability tending to $1$ as $n \rightarrow \infty$, the equation
$S_{n} (\mathbold{\theta})=0$ has a unique solution,
say $\widehat{\boldsymbol{\theta}}$. Furthermore, $\widehat{\boldsymbol{\theta}}$ is strongly consistent and asymptotically normal,
\begin{equation*}
\sqrt{n}(\widehat{\boldsymbol{\theta}}-\boldsymbol{\theta}_0) \xrightarrow{\text{d}} N(0, {\bf H}^{-1} {\bf G} {\bf H}^{-1})
\end{equation*}
where the matrices ${\bf G}(\mathbold{\theta})$ and ${\bf H}(\mathbold{\theta})$ are defined by
\begin{eqnarray}
{\bf G}(\mathbold{\theta}) = \mbox{E} \Biggl[ \frac{ \partial {\mathbold{\lambda}}^{T}_{t}({\mathbold{\theta}})}{ \partial {\mathbold{\theta}}} {\bf D}^{-1}_{t}(\mathbold{\theta})
 \mathbold{\Sigma}_{t}(\mathbold{\theta})
{\bf D}^{-1}_{t}(\mathbold{\theta}) \frac{ \partial {\mathbold{\lambda}}_{t}(\mathbold{\theta})}{ \partial {\mathbold{\theta}^{T}}} \Biggr],
\label{information limit}
\end{eqnarray}
\begin{eqnarray}
{\bf H}(\mathbold{\theta})= \mbox{E} \Biggl[ \frac{ \partial {\mathbold{\lambda}}^{T}_{t}(\mathbold{\theta})}{ \partial {\mathbold{\theta}}} {\bf D}^{-1}_{t}(\mathbold{\theta})
\frac{ \partial {\mathbold{\lambda}}_{t}(\mathbold{\theta})}{ \partial {\mathbold{\theta}^{T}}} \Biggr]
\label{Hessian limit}
\end{eqnarray}
and expectation is taken with respect to the stationary distribution of $\{\mathbold{Y}_{t}\}$.
\label{Theorem1}
\end{Theorem}

When the components of the time series $\{ \mathbold{Y}_{t}$\} are uncorrelated,
then ${\mathbold{\Sigma}}_{t}= {\mathbold D}_{t}$ and therefore the matrices
${\mathbold G}$ and ${\mathbold H}$ coincide. Hence, we obtain a  standard result for the ordinary MLE
in this case. All the above quantities can be calculated by their respective sample counterparts.

\begin{Remark}
\normalfont
To conclude the proof of Theorem \ref{Theorem1} we need to show that the expected value of all
third derivatives of the log-likelihood function \eqref{quasi-loglikelihood} of the perturbed model \eqref{perturbed linear}
within the neighborhood of the true parameter $O(\mathbold{\theta}_{0})$ are uniformly bounded. Additionally, we need to
show that the all third derivatives of the unperturbed model \eqref{linear model in terms of Poisson} are "close" to the
third derivatives of \eqref{perturbed linear}. This point was documented in several publications including \citet{Fokianosetal(2009)} (for
the case of linear model) and \citet{FokianosandTjostheim(2011)} (for the case of the log-linear model). In the appendix, we outline
the methodology of obtaining this result.
\label{remark thirdderivativemqle}
\end{Remark}

For completeness of presentation, we consider briefly QMLE inference for the case of the log-linear model \eqref{loglinear model in terms of Poisson}.
Given the log-likelihood function \eqref{quasi-loglikelihood} we obtain the score, Hessian matrix and conditional information matrix by
\begin{eqnarray}
S_{n}(\mathbold{\theta})= \sum_{t=1}^{n} \sum_{i=1}^{p} \Bigl( y_{i,t}- \exp(\nu_{i,t}(\mathbold{\theta})) \Bigr)
\frac{ \partial \nu_{i,t} (\mathbold{\theta})}{\partial \mathbold{\theta}}=
\displaystyle \sum_{t=1}^{n} \frac{ \partial {\mathbold{\nu}}^{T}_{t}(\mathbold{\theta})}{ \partial {\mathbold{\theta}}}
\Bigl({\bf Y}_{t}- \exp({\mathbold{\nu}}_{t}({\mathbold{\theta}})\Bigr)  ,
\label{scoreloglinear}
\end{eqnarray}
\begin{eqnarray*}
\mathbold{H}_{n}(\mathbold{\theta}) = \sum_{t=1}^{n} \sum_{i=1}^{p} \exp(\nu_{i,t}(\mathbold{\theta}))
\frac{ \partial \nu_{i,t} (\mathbold{\theta})}{\partial \mathbold{\theta}}\frac{ \partial \nu_{i,t} (\mathbold{\theta})}{\partial \mathbold{\theta}^{T}}
-
\sum_{t=1}^{n} \sum_{i=1}^{p} \Bigl( y_{i,t}- \exp(\nu_{i,t}(\mathbold{\theta})) \Bigr)
\frac{ \partial^{2} \nu_{i,t} (\mathbold{\theta})}{\partial \mathbold{\theta}\partial{\mathbold{\theta}^{T}}},
\end{eqnarray*}
\begin{eqnarray*}
\mathbold{G}_{n}( \mathbold{\theta}) = \sum_{t=1}^{n} \sum_{i=1}^{p} \exp(\nu_{i,t}(\mathbold{\theta}))
\frac{ \partial \nu_{i,t} (\mathbold{\theta})}{\partial \mathbold{\theta}}\frac{ \partial \nu_{i,t} (\mathbold{\theta})}{\partial \mathbold{\theta}^{T}},
\end{eqnarray*}
respectively. The recursions  for $\partial \mathbold{\nu}_{t}(\mathbold{\theta})/ \partial \mathbold{\theta}^{T}$
required for computing the QMLE   are obtained  as in  \eqref{recursions linear} but with $\mathbold{\lambda}_{t}$ replaced
by $\mathbold{\nu}_{t}$ and $\mathbold{Y}_{t-1}$ by $\log(\mathbold{Y}_{t-1}+{\bf 1}_{p})$. In summary, we have the following result;
its proof is omitted since it uses identical arguments as those in the proof of Theorem \ref{Theorem1}. Note however  that one of the main ingredients
of the proof is to show that the score function \eqref{scoreloglinear} is a square integrable martingale; this  fact is guaranteed by the conclusions of
Lemma \ref{lemma approx loglinear}; in particular the fourth result.

\begin{Theorem}
\normalfont
Consider model \eqref{loglinear model in terms of Poisson}. Let $\mathbold{\theta} \in \boldsymbol{\Theta} \subset \mathbb{R}^{d}$.
Suppose that $\boldsymbol{\Theta}$ is compact and assume that
 the true value $\mathbold{\theta}_{0}$ belongs to the interior of $\boldsymbol{\Theta}$.
Suppose that at the true value $\mathbold{\theta}_0$, the conditions of Proposition \ref{ergodicity loglinear perturbed} hold true. 
Then there exists a fixed open neighborhood, say $O(\mathbold{\theta}_0)$, of $\mathbold{\theta}_0$
such that with probability tending to $1$ as $n \rightarrow \infty$, the equation
$S_{n} (\mathbold{\theta})=0$, where $S_{n}(\cdot)$ is defined by \eqref{scoreloglinear}, has a unique solution,
say $\widehat{\boldsymbol{\theta}}$. Furthermore, $\widehat{\boldsymbol{\theta}}$ is strongly consistent and asymptotically normal,
\begin{equation*}
\sqrt{n}(\widehat{\boldsymbol{\theta}}-\boldsymbol{\theta}_0) \xrightarrow{\text{d}} N(0, {\bf H}^{-1} {\bf G} {\bf H}^{-1})
\end{equation*}
where the matrices ${\bf G}(\mathbold{\theta})$ and ${\bf H}(\mathbold{\theta})$ are defined by
\begin{eqnarray*}
{\bf G}(\mathbold{\theta}) = \mbox{E} \Biggl[ \frac{ \partial {\mathbold{\nu}}^{T}_{t}(\mathbold{\theta})}{ \partial {\mathbold{\theta}}}
 \mathbold{\Sigma}_{t}(\mathbold{\theta})
\frac{ \partial {\mathbold{\nu}}_{t}(\mathbold{\theta})}{ \partial {\mathbold{\theta}^{T}}} \Biggr],  \quad
{\bf H}(\mathbold{\theta})= \mbox{E} \Biggl[ \frac{ \partial {\mathbold{\nu}}^{T}_{t}(\mathbold{\theta})}{ \partial {\mathbold{\theta}}} {\bf D}_{t}(\mathbold{\theta})
\frac{ \partial {\mathbold{\nu}}_{t}(\mathbold{\theta})}{ \partial {\mathbold{\theta}^{T}}} \Biggr]
\end{eqnarray*}
and expectation is taken with respect to the stationary distribution of $\{\mathbold{Y}_{t}\}$.
\label{Theorem2}
\end{Theorem}

\noindent
Although the product form of \eqref{quasi-loglikelihood} indicates independence, the dependence structure in \eqref{linear model in terms of Poisson}
and \eqref{loglinear model in terms of Poisson} will be picked up explicitly through the dependence of \eqref{quasi-loglikelihood}
on the matrices $\mathbold{A}$ and $\mathbold{B}$. The copula structure, however, does not explicitly appear in \eqref{quasi-loglikelihood}, even though indirectly it does
because of the conditional innovation  $\mathbold{Y}_{t} \mid \mathbold{\lambda}_{t}$. (One could, of course, have chosen a more specific dependence model
for these quantities. The copula was chosen because of its general way of describing dependence.)
To recover the copula dependence one has to look at the conditional
distribution of $\mathbold{Y}_{t} \mid \mathbold{\lambda}_{t}$ and compare it with the conditional distribution of $\mathbold{Y}^{*}_{t} \mid \mathbold{\lambda}_{t}$, say,
generated by a suitable copula model conditional on $\mathbold{\lambda}_{t}$. There are several ways of comparing such distributions, e.g. the Kullback-Leibler or
Hellinger distances. A thorough study of this problem requires a separate publication. In the appendix \ref{sec:copula estimation},
we have opted for a preliminary and heuristic approach based on the newly developed concept of local Gaussian correlation (see Appendix \ref{Sec:LGC}).

\section{Simulation and data analysis}
\label{sec:examples}

In this section we illustrate the theory by presenting a limited simulation study for both  linear and log-linear models. In addition we
include a  real data example. Maximum likelihood estimators are
calculated by optimization of the log-likelihood function \eqref{quasi-loglikelihood}.

\subsection{Simulations for the multivariate linear model}

For the simulation study we only consider a  two-dimensional process, that is  $p=2$.
To initiate the maximization  algorithm,  we obtain  starting values for the parameter
vector $\mathbold{\theta} = (\mathbf{d}, \vect^{T}(\mathbf{A}), \vect^{T}(\mathbf{B}))$
by a linear regression fit to the data; see \citet{Fokianosetal(2009)} and \citet{Fokianos(2015)} for the analogous method in the univariate
case. Throughout the simulations we generate 1000 realizations with sample sizes of 500 and 1000.
We report the estimates of the parameters by averaging out the results from all simulations, and similarly, the standard errors
correspond to the sampling standard errors of the estimates obtained by the simulation.
Table  \ref{tab:1} lists  the results obtained for \eqref{linear model mult} by using  

\begin{eqnarray}
{\mathbf A} =
\begin{pmatrix}
0.3 & 0\\
0 & 0.25
\end{pmatrix},
\quad
{\mathbf B} = \begin{pmatrix}
0.5 & 0\\
0 & 0.4
\end{pmatrix}
~\mbox{and}~
{\mathbf d} = (1, 2).
\label{first linear specification}
\end{eqnarray}
To generate the data, we employ  the Gaussian copula  with parameter $\phi$ chosen as
$0$ and 0.5. Obviously the case $\phi=0$ corresponds  to a two-dimensional process with independent components.
Table \ref{tab:1} illustrates  that the estimated parameters approach their true values  quite adequately
while their  the standard errors are relatively small and in line with previous studies.
Figure \ref{fig:1}  supports further the asymptotic normality of the estimators.

\begin{table}[htbp] \small
 \centering
 \begin{tabular}{c|c|cccccccccc}\hline \hline
Sample size & $\phi$  & $\hat{d}_1$ & $\hat{d}_2$ & $\hat{a}_{11}$  & $\hat{a}_{22}$  &$\hat{b}_{11}$ & $\hat{b}_{22}$ & $\hat{a}_{12}$  & $\hat{a}_{21}$  &$\hat{b}_{12}$ & $\hat{b}_{21}$ \\ \hline
\multirow{4}{*}{500} & \multirow{2}{*} 0 &  1.056 & 2.153& 0.294& 0.236& 0.495& 0.394& 0.034& 0.030 &0.018& 0.020 \\
 & &  (0.206) & (0.432)& (0.072)& (0.099) &(0.050)& (0.047)& (0.052)& (0.045) &(0.026)& (0.029) \\ \cline{2-12}
& \multirow{2}{*} {0.5} &  1.074& 2.145& 0.292& 0.239& 0.495& 0.395& 0.042& 0.036& 0.018& 0.020   \\
&  & (0.211)& (0.416)& (0.081)& (0.106)& (0.053)& (0.051) &(0.061)& (0.052)& (0.027)& (0.030)        \\ \hline
\multirow{4}{*}{1000} & \multirow{2}{*} 0 & 1.035& 2.083& 0.299& 0.241& 0.495& 0.398& -0.001&  $-2 \cdot 10^{-4}$ &0.001&$ -2 \cdot 10^{-4}$   \\
 &&  (0.152)& (0.314)& (0.053)& (0.072)& (0.035)& (0.033)& (0.064)& (0.052)& (0.030) &(0.034)  \\ \cline{2-12}
 & \multirow{2}{*} {0.5} & 1.045 &2.059& 0.294& 0.247 &0.495& 0.396& -0.001& -0.001& 0.001 &$ 3 \cdot 10^{-4}$ \\
 && (0.149) &(0.294)& (0.056)& (0.074)& (0.038)& (0.037)& (0.072)& (0.056)& (0.033)& (0.037) \\ \hline
\end{tabular}
\caption{Simulation results for the multivariate linear model \eqref{linear model mult} by employing
the Gaussian copula with parameter $\phi$. True parameter values are given by \eqref{first linear specification}.
Standard errors of the estimators  are given in parentheses. Results are based on 1000 runs.}
\label{tab:1}
\end{table}

\normalsize

\begin{figure}
\centering
 \includegraphics[scale=0.5]{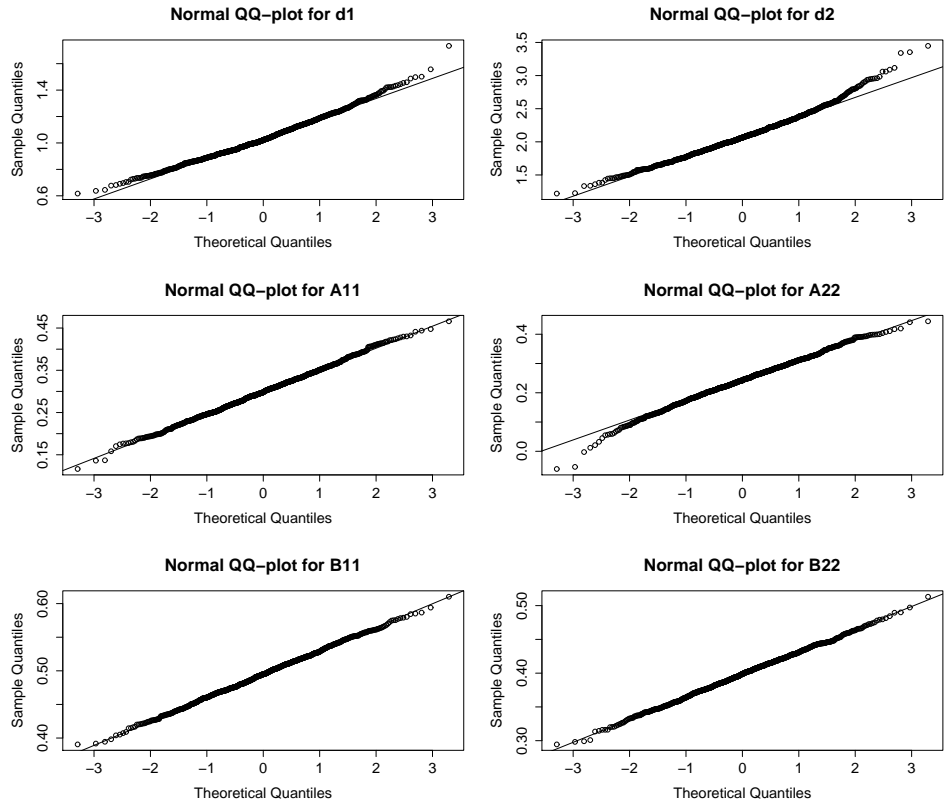}
 \caption{QQ-plots of the standardized sampling distribution of $\widehat{\mathbold{\theta}}$ for the multivariate linear model with true parameter values  given
  by \eqref{first linear specification} with $n=1000$. Data have been generated by a   Gaussian copula with $\phi=0$. Results are based on 1000 runs. }
\label{fig:1}
\end{figure}

Furthermore, we study the sensitivity of the previous results on  the choice of  the copula function employed to generate data.
This is done by   a further  simulation setup  which utilizes   the Clayton copula with parameters $\phi = 0$ and 1.
The parameter vector $\mathbold{\theta}$ is chosen  according to \eqref{first linear specification} and  the results of this study
are reported  in Table \ref{tab:3} and Figure \ref{fig:2}  both of which indicate the adequacy of the proposed estimation method.

\begin{table}[htbp] \small
 \centering
 \begin{tabular}{c|c|cccccccccc}\hline \hline
Sample size & $\phi$  & $\hat{d}_1$ & $\hat{d}_2$ & $\hat{a}_{11}$  & $\hat{a}_{22}$  &$\hat{b}_{11}$ & $\hat{b}_{22}$ & $\hat{a}_{12}$  & $\hat{a}_{21}$  &$\hat{b}_{12}$ & $\hat{b}_{21}$ \\ \hline
\multirow{4}{*}{500} & \multirow{2}{*} 0 & 1.082 & 2.173& 0.290& 0.235 &0.495& 0.394 &0.002& -0.004& 0.001& 0.002  \\
 & &  (0.220)& (0.423)& (0.077)& (0.104)& (0.049)& (0.050)& (0.093) &(0.076) &(0.044) &(0.049) \\ \cline{2-12}
& \multirow{2}{*} {1} & 1.076& 2.156 &0.289& 0.243& 0.494& 0.394& 0.002& -0.006&   $1 \cdot 10^{-4}$  &-0.001   \\
&  &  (0.215)& (0.413)& (0.087)& (0.112)& (0.058)& (0.054)& (0.106)& (0.089)& (0.049)& (0.057)       \\ \hline
\multirow{4}{*}{1000} & \multirow{2}{*} 0 & 1.032 &2.058& 0.298& 0.247& 0.496& 0.398& -0.001& -0.001 &   $-2 \cdot 10^{-4}$ &   $-1 \cdot 10^{-4}$  \\
 &&   (0.145)& (0.299)& (0.051)& (0.070)& (0.034)& (0.034)& (0.065)& (0.056)& (0.029)& (0.035)   \\ \cline{2-12}
 & \multirow{2}{*} {1} & 1.045 &2.077& 0.293& 0.247& 0.497& 0.397& 0.001& -0.005&  $-3 \cdot 10^{-4}$& 0.001 \\
 &&  (0.151)& (0.286)& (0.059)& (0.078)& (0.040)& (0.038)& (0.074)& (0.064)& (0.035)& (0.041) \\ \hline
\end{tabular}
\caption{Simulation results for the multivariate linear model \eqref{linear model mult} by employing
the Clayton  copula  with parameter $\phi$. True parameter values are given by \eqref{first linear specification}. Standard errors of the estimators  are given in parentheses.
Results are based on 1000 runs.}
\label{tab:3}
\end{table}

\begin{figure}
\centering
 \includegraphics[scale=0.5]{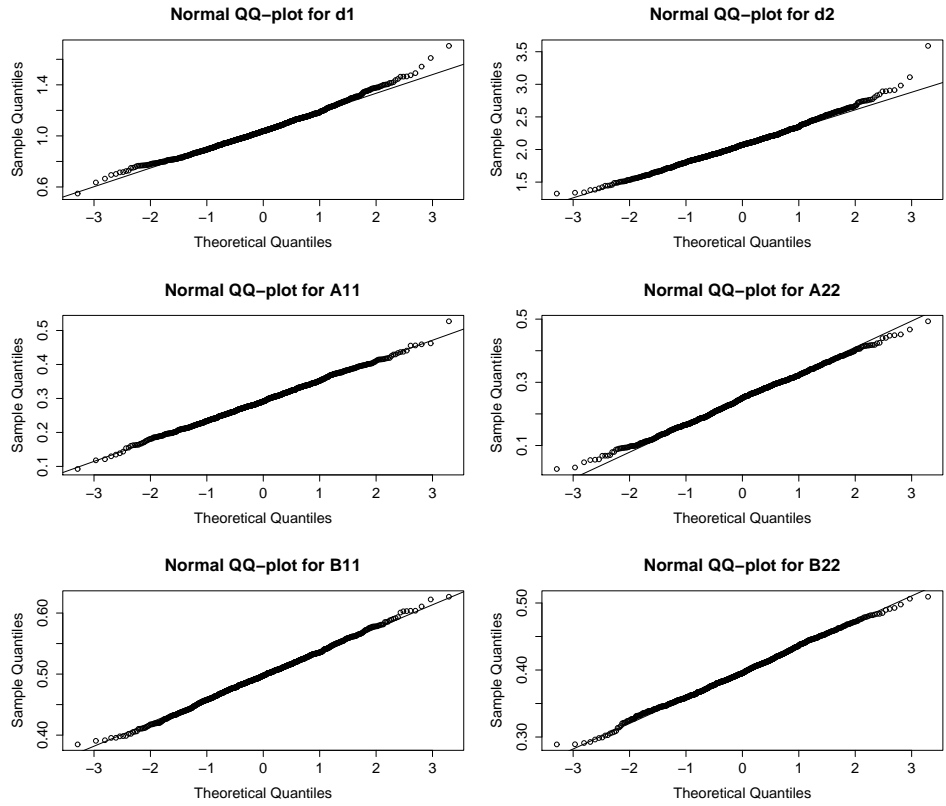}
 \caption{QQ-plots of the standardized sampling distribution of $\widehat{\mathbold{\theta}}$ for the multivariate linear model with true parameter values  given
 by \eqref{first linear specification} with $n=1000$. Data have been generated by a  Clayton  copula with $\phi=0$. Results are based on 1000 runs.}
\label{fig:2}
\end{figure}

Finally Table  \ref{tab:2} illustrates   simulation  results obtained from the linear model where the off-diagonal elements of the matrices
${\mathbf A}$ and ${\mathbf B}$ are non-zero, i.e.
following parameters
\begin{eqnarray}
{\mathbf A} = \begin{pmatrix}
0.3 & 0.05\\
0.1 & 0.25
\end{pmatrix},
\quad
{\mathbf B} = \begin{pmatrix}
0.5 & 0.05\\
0.1 & 0.4
\end{pmatrix}
~\mbox{and}~
{\mathbf d} = (0.5, 1).
\label{second linear specification}
\end{eqnarray}
Note that these parameter values yield $\|| {\bf A}+  {\bf B} \||_{2} =0.89 <1$ but $\|| {\bf A}\||_{1}+  \||{\bf B} \||_{1}=1$ (compare Propositions
\ref{ergodicity linear perturbed} and \ref{ergodicity linear perturbed wd}).

\begin{table}[htbp] \small
 \centering
 \begin{tabular}{c|c|cccccccccc}\hline \hline
Sample size & $\phi$  & $\hat{d}_1$ & $\hat{d}_2$ & $\hat{a}_{11}$  & $\hat{a}_{22}$  &$\hat{b}_{11}$ & $\hat{b}_{22}$ & $\hat{a}_{12}$  & $\hat{a}_{21}$  &$\hat{b}_{12}$ & $\hat{b}_{21}$ \\ \hline
\multirow{4}{*}{500} & \multirow{2}{*} 0 & 0.871 &1.421 &0.289 &0.222 &0.493 &0.396 &0.087 &0.167 &0.051& 0.098  \\
 & & (0.205)& (0.349)& (0.071)& (0.084)& (0.049)& (0.050) &(0.082) &(0.077)& (0.045)& (0.049)  \\ \cline{2-12}
& \multirow{2}{*} {0.5} &  0.772 &1.116& 0.279& 0.200& 0.494& 0.395& 0.083& 0.161& 0.051& 0.099  \\
&  &    (0.170)& (0.264)& (0.074)& (0.087)& (0.051)& (0.051)& (0.085)& (0.081)& (0.050)& (0.052)      \\ \hline
\multirow{4}{*}{1000} & \multirow{2}{*} 0 & 0.803& 1.316 &0.295 &0.222 &0.498 &0.400 &0.083 &0.166 &0.052& 0.099    \\
 &&   (0.134)& (0.236)& (0.052)& (0.057)& (0.036)& (0.032)& (0.054)& (0.054)& (0.030) &(0.036)   \\ \cline{2-12}
 & \multirow{2}{*} {0.5} & 0.733 &1.056 &0.286 &0.207 &0.497 &0.396& 0.082& 0.157 &0.048 &0.100 \\
 &&  (0.118)& (0.181)& (0.055)& (0.061)& (0.037)& (0.037)& (0.057)& (0.054)& (0.035)& (0.037)  \\ \hline
\end{tabular}
\caption{Simulation results for the multivariate linear model \eqref{linear model mult} by employing
the Clayton  copula  with parameter $\phi$. True parameter values are given by \eqref{second linear specification}. Standard errors of the estimators  are given in parentheses.
Results are based on 1000 runs.}
\label{tab:2}
\end{table}

\subsection{Simulations for the log-linear model}

In this section, we report some limited simulation study results for  the case of the log-linear model \eqref{loglinear model in terms of Poisson}.
For this model, the problem of obtaining starting values for the parameter $\mathbold{\theta}$ to initiate the maximization of \eqref{quasi-loglikelihood}
is more challenging. We resort to univariate fits (see \citet{FokianosandTjostheim(2011)} for details) in the case of diagonal
matrices $\mathbf{A}$ and $\mathbf{B}$. In the case of non-diagonal matrices, we can still fit univariate log-linear models
to each series and then add an extra step of multivariate least squares estimation to obtain starting values.
The parameter values have been chosen by
\begin{eqnarray}
{\mathbf A} = \begin{pmatrix}
-0.3 & 0\\
0 & 0.25
\end{pmatrix},
\quad
{\mathbf B} = \begin{pmatrix}
0.5 & 0\\
0 & 0.4
\end{pmatrix},~~
\mathbf{d} = (0.5, 1),
\label{first log-linear specification}
\end{eqnarray}
and
\begin{eqnarray}
{\mathbf A} = \begin{pmatrix}
0.4 & 0\\
0 & 0.45
\end{pmatrix},
\quad
{\mathbf B} = \begin{pmatrix}
0.5 & 0\\
0 & 0.35
\end{pmatrix},~~
{\mathbf d} = (0.3, 0.5).
\label{second log-linear specification}
\end{eqnarray}
Furthermore, we employ the Gaussian copula  to generate the data
with chosen  parameter $\phi$ given by  0 and 0.5. The simulation results are shown  in Tables \ref{tab:4} and \ref{tab:5}.
In both cases we note that the  estimated parameters are close to their true values, and the approximation improves
for larger sample sizes. QQ-plots for the standardized
estimated parameters under model  \eqref{first log-linear specification} when  $\phi=0.5$ and  sample size of 1000 observations,
are shown  in Figure \ref{fig:3}, and again support asymptotic normality.

\begin{table}[htbp] \small
 \centering
 \begin{tabular}{c|c|cccccccccc}\hline \hline
Sample size & $\phi$  & $\hat{d}_1$ & $\hat{d}_2$ & $\hat{a}_{11}$  & $\hat{a}_{22}$  &$\hat{b}_{11}$ & $\hat{b}_{22}$ & $\hat{a}_{12}$  & $\hat{a}_{21}$  &$\hat{b}_{12}$ & $\hat{b}_{21}$ \\ \hline
\multirow{4}{*}{500} & \multirow{2}{*} 0 &  0.510& 1.042& -0.302& 0.238& 0.498& 0.3986&  $5 \cdot 10^{-4}$ &0.011 &  $-1 \cdot 10^{-4}$ &0.002  \\
 & & (0.661)& (0.248)& (0.121)& (0.107)& (0.064)& (0.046)& (0.045)& (0.268)& (0.021)& (0.127)  \\ \cline{2-12}
& \multirow{2}{*} {0.5} &  0.533 &1.027& -0.310 &0.242& 0.498& 0.399& -0.002& 0.007&    $2 \cdot 10^{-4}$ &0.002  \\
&  &   (0.699)& (0.238)& (0.118)& (0.105)& (0.068)& (0.047)& (0.045)& (0.280)& (0.022)& (0.131)      \\ \hline
\multirow{4}{*}{1000} & \multirow{2}{*} 0 &  0.504 &1.016& -0.303& 0.246& 0.496& 0.399&    $2 \cdot 10^{-4}$& -0.001 &   $3 \cdot 10^{-4}$ &0.004   \\
 && (0.515)& (0.163)& (0.086)& (0.071)& (0.045)& (0.034)& (0.032)& (0.196)& (0.016)& (0.089)    \\ \cline{2-12}
 & \multirow{2}{*} {0.5} & 0.536& 1.005 &-0.299 &0.250 &0.498 &0.399 &-0.001 &-0.010 &0.001 &-0.001  \\
 &&  (0.502) &(0.166)& (0.091)& (0.071)& (0.045)& (0.032)& (0.031)& (0.196)& (0.015) &(0.091) \\ \hline
\end{tabular}
\caption{Simulation results for the multivariate log-linear model \eqref{log-linear model mult} by employing
the Gaussian   copula with parameter $\phi$. True parameter values are given by \eqref{first log-linear specification}. Standard errors are given in parentheses. Results are based on 1000 runs.}
\label{tab:4}
\end{table}

\begin{figure}
\centering
 \includegraphics[scale=0.5]{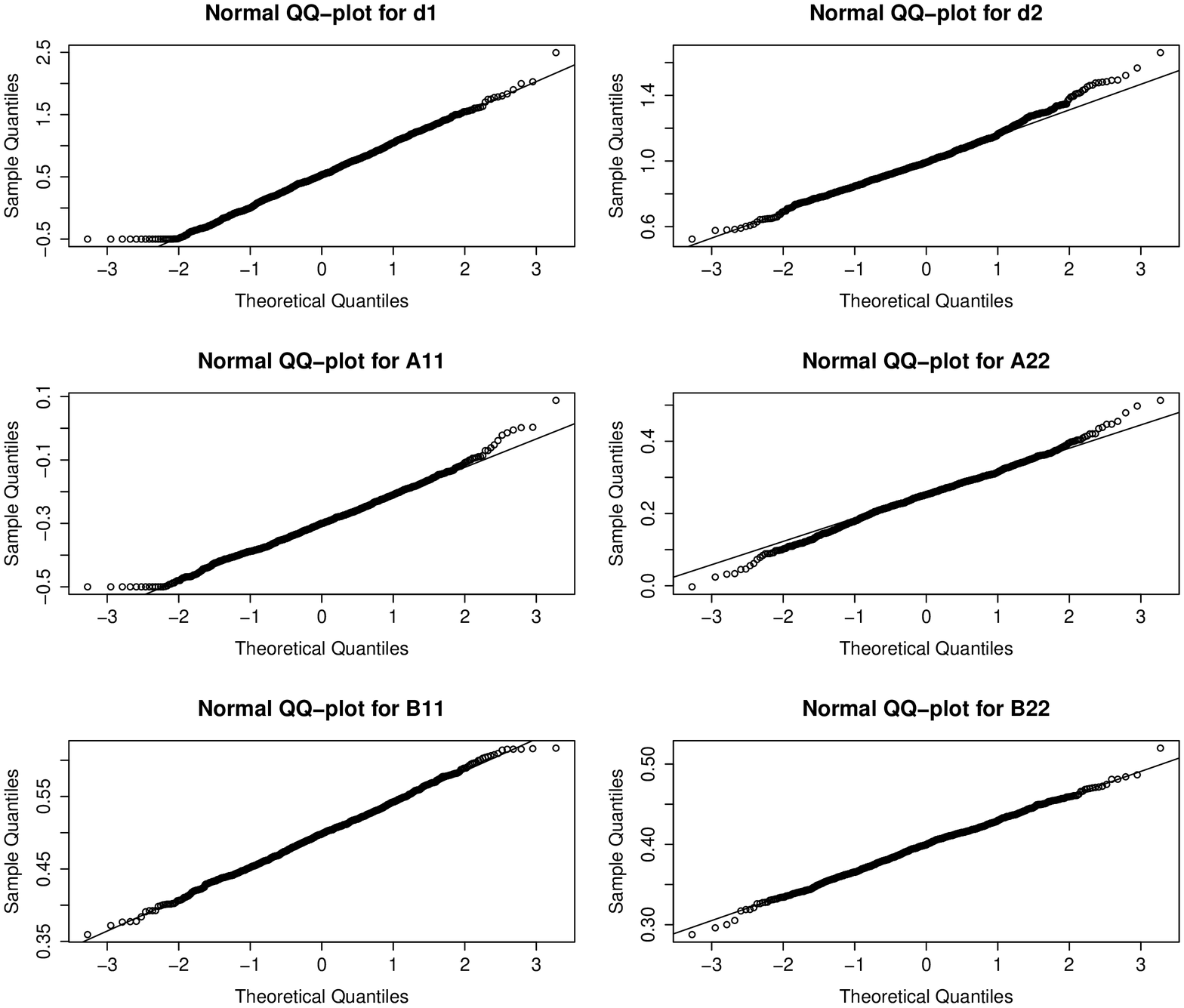}
 \caption{QQ-plots of the standardized sampling distribution of $\widehat{\mathbold{\theta}}$ for the multivariate  log-linear model with  true parameter values   given
  by \eqref{first log-linear specification} with $n=1000$. Data have been generated by a  Gaussian   copula with $\phi=0.5$. Results are based on 1000 runs.}
\label{fig:3}
\end{figure}

\begin{table}[htbp] \small
 \centering
 \begin{tabular}{c|c|cccccccccc}\hline \hline
Sample size & $\phi$  & $\hat{d}_1$ & $\hat{d}_2$ & $\hat{a}_{11}$  & $\hat{a}_{22}$  &$\hat{b}_{11}$ & $\hat{b}_{22}$ & $\hat{a}_{12}$  & $\hat{a}_{21}$  &$\hat{b}_{12}$ & $\hat{b}_{21}$ \\ \hline
\multirow{4}{*}{500} & \multirow{2}{*} 0 &  0.321 & 0.522 &0.384 &0.426& 0.505 & 0.353& 0.011 & 0.006&  $-4 \cdot 10^{-4} $ & $-3 \cdot 10 ^{-4}$   \\
 & & (0.183)& (0.244)& (0.064)& (0.096)& (0.050)& (0.047)& (0.076)& (0.068)& (0.058)& (0.034)  \\ \cline{2-12}
& \multirow{2}{*} {0.5} &  0.326 &0.524 &0.388 &0.428 & 0.500 &0.352 &0.013& 0.004  &-0.004  & $ -1 \cdot 10^{-4}$  \\
&  &   (0.152) &(0.198) &(0.066) & (0.109) & (0.050) & (0.051) &(0.084) & (0.074) & (0.064)  &(0.037)     \\ \hline
\multirow{4}{*}{1000} & \multirow{2}{*} 0 & 0.302& 0.518& 0.394 &0.437& 0.501& 0.350& 0.007& 0.006& -0.002 &-0.001     \\
 && (0.129)& (0.158)& (0.043)& (0.065)& (0.033)& (0.035)& (0.055)& (0.048)& (0.041)& (0.024)    \\ \cline{2-12}
 & \multirow{2}{*} {0.5} &  0.321 &0.521& 0.394& 0.436& 0.502& 0.350& 0.003& -0.003& 0.002& 0.001\\
 &&  (0.109) &(0.132)& (0.046)& (0.068)& (0.035)& (0.035)& (0.057)& (0.052)& (0.045) &(0.027) \\ \hline
\end{tabular}
\caption{Simulation results for the multivariate log-linear model \eqref{log-linear model mult} by employing
the Gaussian   copula with parameter $\phi$. True parameter values are given by \eqref{second log-linear specification}. Standard errors are given in parentheses. Results are based on 1000 runs.}
\label{tab:5}
\end{table}

\noindent
To evaluate the proposed procedure for the copula parameter estimation (see Section \ref{sec:copula estimation})
we perform a further simulation study for the linear model \eqref{linear model mult} using  the
parameter values defined by  \eqref{first linear specification}. The results are reported in Appendix \ref{sec:simresult:copula}. (Note that for the 
linear and log-linear model we do not need the copula structure to estimate the parameters $\mathbold{d}$, $\mathbold{A}$
and $\mathbold{B}$.)

\subsection{Real data analysis}

We fit the linear and log-linear models to a bivariate count time series
which consists of  the number of transactions per 15 seconds for the stocks Coca-Cola Company (KO) and IBM on September 19th 2005.
There are 1551 observations in each of the two series, covering trades from 09:30 to 16:30, excluding the first and last minute of transactions.
Figure \ref{fig:predb} shows a time series plot of the  data and Figure \ref{fig:auto} depicts
the autocorrelation function and cross- autocorrelation functions. Clearly, the plot of the autocorrelation functions reveals high correlation  within and between the individual transaction series. Note further that
mean number of transactions  is  5.115 and 4.470 , for  IBM and KO stocks respectively.
The sample variances  are 17.315 (IBM) and 12.806 (KO),
that is the data clearly shows marginal overdispersion.

\begin{figure}[htb]
\centering
 \includegraphics[scale=0.45]{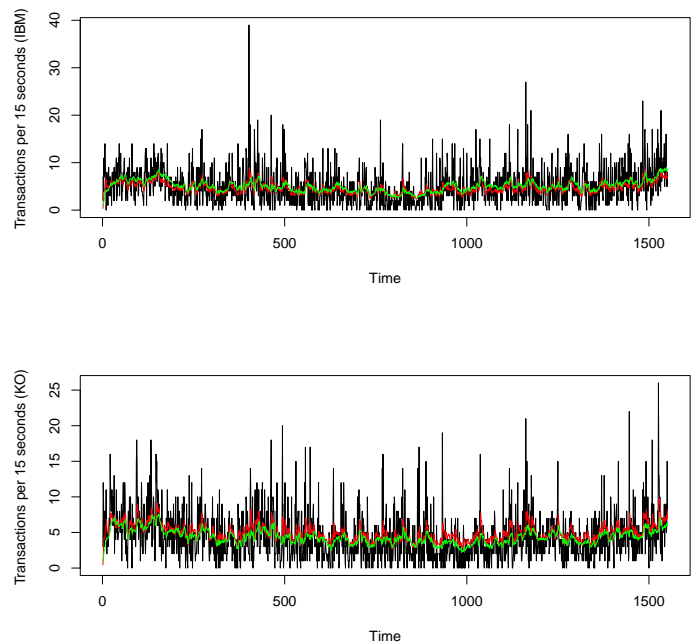}
 \caption{Number of transactions per 15 seconds for IBM (top) and Coca-Cola (bottom) and the respective predicted number of transactions from the linear model (red line) and the log-linear model (green line).}
\label{fig:predb}
\end{figure}

\begin{figure}[htb]
 \centering
 \includegraphics[scale=0.40]{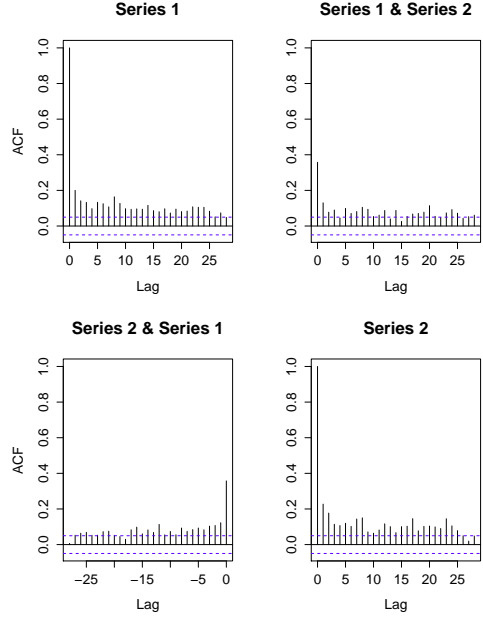}
 \caption{Auto- and cross-correlation function of the transaction data.}
\label{fig:auto}
\end{figure}

Maximization of the quasi log-likelihood function \eqref{quasi-loglikelihood}, where we have initialized the recursions by a linear regression
fit to the data, as it was done for  the simulation experiments, yields   the following results:
\begin{eqnarray*}
\hat{\lambda}_{1,t} & = &  0.3335 + 0.6993 \hat{\lambda}_{1,t-1} + 0.1230 \hat{\lambda}_{2, t-1}+ 0.2027 Y_{1, t-1}+ 0.0150 Y_{2, t-1}, \\
&&(0.0074) \;\;\;  (0.0649) \;\;\;\;\;\;\;\; (0.0007)\;\;\;\;\;\;\;\; (0.2024) \;\;\;\;\;\;\;\; (0.0046) \\
\hat{\lambda}_{2,t} & = &  0.4308 + 0.0459 \hat{\lambda}_{1,t-1} + 0.0527 \hat{\lambda}_{2, t-1}+ 0.1594 Y_{1, t-1}+ 0.5635 Y_{2, t-1}\\
&&(0.4289)\;\;\; (0.0037) \;\;\;\;\;\;\;\; (0.0005)\;\;\;\;\;\;\;\; (0.0461) \;\;\;\;\;\;\;\; (0.2756)
\end{eqnarray*}
For fitting the log-linear model  initialization of the recursions has been done as in Section \ref{sec:examples} when considering non-diagonal matrices.
The results are as follows:
\begin{eqnarray*}
\hat{\nu}_{1,t} & = &  0.0416 + 0.9025 \hat{\nu}_{1,t-1} - 0.0264 \hat{\nu}_{2, t-1}+ 0.0751 \log (Y_{1, t-1}+1) + 0.0317 \log(Y_{2, t-1}+1), \\
&&(2.65 \cdot 10^{-5}) \;\;\;  (4.37\cdot 10^{-5}) \;\; (1.30\cdot 10^{-5})\;\;(4.16\cdot 10^{-5}) \;\;\;\;\;\;\;\; (3.74\cdot 10^{-5}) \\
\hat{\nu}_{2,t} & = &  0.0491 - 0.0362 \hat{\nu}_{1,t-1} + 0.8778 \hat{\nu}_{2, t-1}+ 0.0217 \log(Y_{1, t-1} +1) + 0.0914 \log(Y_{2, t-1}+1)\\
&&(2.65 \cdot 10^{-4})\;\;\; (3.05 \cdot 10^{-6}) \;\; (6.70 \cdot 10^{-5})\;\; (4.10 \cdot 10^{-4}) \;\;\;\;\;\;\;\; (1.49 \cdot 10^{-3})
\end{eqnarray*}

In both cases, the standard errors given in parentheses underneath  the estimated parameters were computed using the robust estimator of the covariance  matrix, $\mathbf{H}^{-1}_n(\mathbf{\hat{\theta}})\mathbf{G}_n(\mathbf{\hat{\theta}})\mathbf{H}^{-1}_n(\mathbf{\hat{\theta}})$ where $\mathbf{H}_{n}$ and $\mathbf{G}_{n}$ are given in equation \eqref{Hessian}
and \eqref{information}, respectively. The magnitude of the standard errors shows that the feedback process should be considered in both models.
\\
The predictions obtained from both  fitted  models are denoted  by $\hat{Y}_{i, t} = \lambda_{i,t}(\mathbf{\hat{\theta}})$ for $i=1,2$
and are shown  in Figure \ref{fig:predb}.
We see that the predictions approximate the observed processes reasonably well.


To examine the model fit, we consider the Pearson residuals, defined by $e_{i,t} = (Y_{i,t} - \lambda_{i,t})/ \sqrt{\lambda_{i,t}}$ for $i=1, 2$.
Under the correct model, the sequence $e_{i,t}$ is a white noise sequence with constant variance. We substitute $\lambda_{i,t}$ by
$\lambda_{i,t}(\mathbf{\hat{\theta}})$ to obtain $\hat{e}_{i,t}$. We compute the Pearson residuals for both models and examine their cumulative periodograms.
Figure \ref{fig:res} supports the marginal whiteness of the residual process. Finally,  the results of the copula estimation are reported in Appendix \ref{sec:copestimrealdata}.
%
%

\begin{figure}[htb]
\centering
  \begin{tabular}{@{}c@{}}
    \includegraphics[scale=0.45]{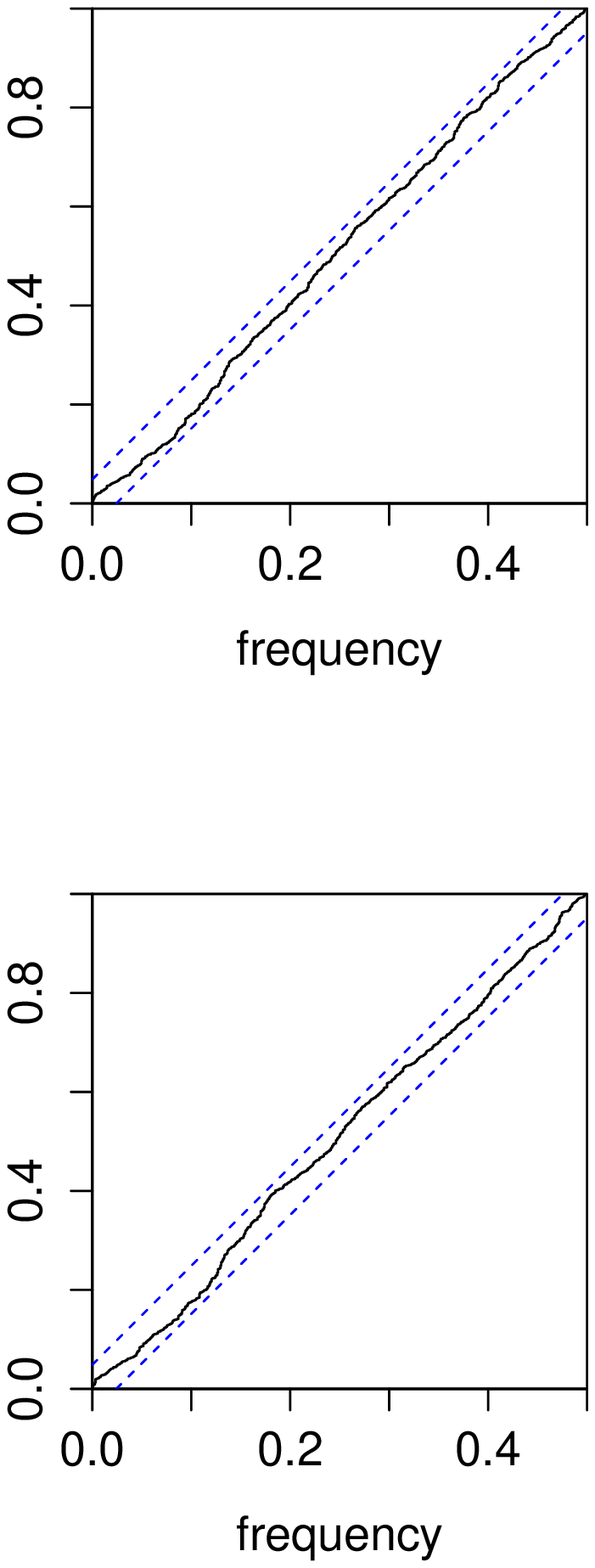}
    \includegraphics[scale=0.45]{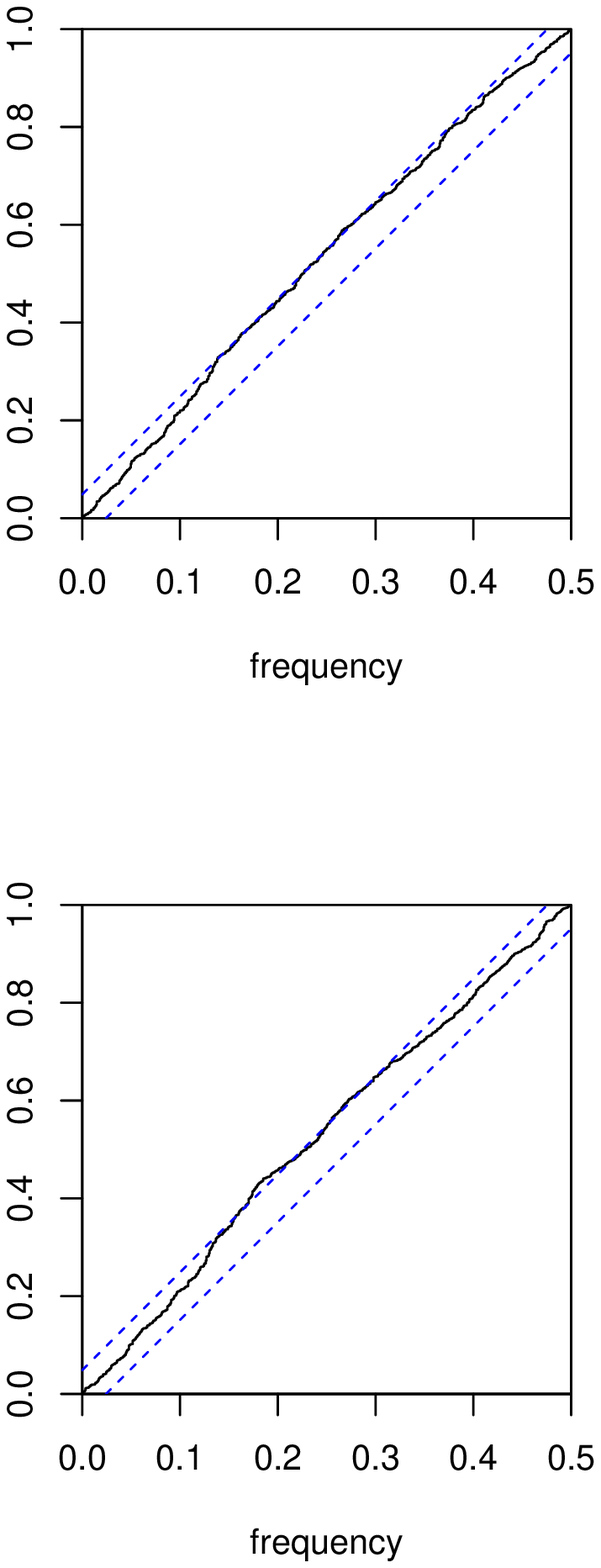}
  \end{tabular}
  \caption{Left:Cumulative periodogram plots of the Pearson residuals from the linear fit of IBM (top) and Coca-Cola (bottom). Right:Cumulative periodogram plots of the Pearson residuals from the log-linear fit of IBM (top) and Coca-Cola (bottom). }
\label{fig:res}
\end{figure}

\section{Discussion}
\label{sec:outlook}

In this work,  we have studied the problem of inference and modeling for multivariate count time series.  We have proposed
models that can accommodate dependence across time and within time series components. Further investigation is
required to develop the properties of these multivariate count processes not the least for the copula estimation.  In addition, equation \eqref{score}
motivates  a more general framework  that can be developed for the analysis of multivariate count time series modes.
For instance, a natural generalization, is to consider
\begin{eqnarray}
S_{{\mathbold v}}({\mathbold \theta}) = \sum_{t=1}^{n} \frac{ \partial {\mathbold{\lambda}}_{t}}{ \partial {\mathbold{\theta}}}
{\bf V}^{-1}_{t}(\rho, {\mathbold \lambda}_{t}({\mathbold \theta}))
\Bigl({\bf Y}_{t}- {\mathbold{\lambda}}_{t}(\mathbold \theta)\Bigr),
\label{score}
\end{eqnarray}
where the notation is completely analogous to the equations \eqref{score independnence} and ${\bf V}^{-1}_{t}(\rho, {\mathbold \lambda}_{t}({\mathbold \theta}))$
is a  $p \times p$ "working" conditional
covariance matrix which depend upon the process $\{ \mathbold{\lambda}_{t} \}$ and possibly some other parameters $\rho$, as we explain below.
Several choices for the working conditional covariance matrix are available in the literature. Here we discuss some of the most commonly used.
\begin{enumerate}
\item ${\bf V}={\bf I}_{p}$. The choice of the identity matrix corresponds to a  least squares solution to the problem of estimating ${\mathbold \theta}$.
\item The $p \times p$ diagonal matrix
$$
{\bf D}({\mathbold \lambda})= \left(
           \begin{array}{cccc}
             \lambda_{1,t} & 0 & \cdots & 0 \\
               0 &  \lambda_{2,t}  & \cdots & 0 \\
             \cdots  &  \cdots & \cdots & \cdots \\
             0 & 0 & \cdots & \lambda_{p,t} \\
           \end{array}
         \right)
$$
yields  estimating equations \eqref{score independnence}; that is the score function under independence.
\item The choice
$$
{\bf V}(\rho, {\mathbold{\lambda}})=\left(       \begin{array}{cccc}
                                                  \lambda_{1, t} & \rho_{12} \sqrt{\lambda_{1,t}}\sqrt{\lambda_{2,t} } & \cdots  & \rho_{1p} \sqrt{\lambda_{1,t}}\sqrt{\lambda_{p,t}}  \\
                                                  \rho_{12} \sqrt{\lambda_{1,t}}\sqrt{\lambda_{2,t}}  &  \lambda_{2,t} &  \cdots  & \rho_{2p} \sqrt{\lambda_{2,t}}\sqrt{\lambda_{p,t}}   \\
                                                  \cdots  &  \cdots  &  \cdots  & \cdots  \\
                                                  \rho_{1p} \sqrt{\lambda_{1,t}}\sqrt{\lambda_{p,t}}   & \rho_{2p} \sqrt{\lambda_{2,t}}\sqrt{\lambda{p,t}} & \cdots & \lambda_{p,t} \\
                                                \end{array}
                                              \right)
$$
yields  a constant conditional correlation type of model for multivariate count time series, see \citet{Terasvirtaetal(2010)}, among others.
\end{enumerate}
We leave this topic for further research by mentioning also  the recent work of  \citet{FrancqandZakoian(2016)} who consider
estimation of  multivariate volatility models equation by equation.

\clearpage

\renewcommand{\theequation}{A-\arabic{equation}}
\renewcommand{\theLemma}{A-\arabic{Lemma}}
\renewcommand{\thesubsection}{A-\arabic{subsection}}
\setcounter{equation}{0}
\setcounter{Lemma}{0}
\section*{Appendix}

It is easy to see that ${\mathbold \lambda}^{\star}=({\bf I}-{\bf A})^{-1} {\bf d}$ is a fixed point of the skeleton \eqref{linear model in terms of Poisson}.
The proof of the following lemma is quite analogous to the proof of \citet[Lemma A.1]{Fokianosetal(2009)} and it is omitted.

\begin{Lemma} \rm
Let $\{ \mathbold{\lambda}_{t}\}$ be a Markov chain defined by \eqref{loglinear model in terms of Poisson} or   \eqref{perturbed linear}. If
$\|| {\bf A}  \||_{2} <1 $, then every point in $[\lambda^{\star}_{1}, \infty)\times \ldots\times[\lambda^{\star}_{p}, \infty)$ is reachable, where
$\lambda^{\star}_{i}$ denotes the $i$'th component of the vector ${\mathbold \lambda}^{\star}$.
\end{Lemma}

\subsection{Proof of Proposition \ref{ergodicity linear perturbed}}
The conditions of  $\phi$-irreducibility and the existence of small sets can be proved along the lines of the proof of
\citet[Prop. 2.1]{Fokianosetal(2009)} provided
that $\|| {\bf A} \||_{2} < 1$. As in the proof of that Proposition we use the Tweedie criterion to prove geometric ergodicity.
Define now the test function $V({\bf x})=1+\|{\bf x}\|_{2}^{r}$.
Then, we obtain as $\lambda_{i} \rightarrow \infty$, $i=1,2,\ldots, p$,
\begin{eqnarray*}
\mbox{E} \left[ V( \mathbold{\lambda}_{t}^{m}) | {\mathbold{\lambda}}_{t-1}^{m}= \mathbold{\lambda} \right]& = &
 1+\mbox{E} \left[  \|{\bf d}+{\bf A} {\mathbold \lambda}+ {\bf B}{ \bf Y}_{t-1}^{m} + {\mathbold{\epsilon}}_{t;m}\|_{2}^{r}\right] \\ & \sim &
 \mbox{E} \left[  \|{\bf A} {\mathbold \lambda}+ {\bf B}{ \bf Y}_{t-1}^{m}\|_{2}^2 \right]^\mu,
\end{eqnarray*}
where we assume, without loss of generality, that  $\mu= r/2$,  $r$ a positive integer.
Next,
\begin{eqnarray*}
 \mbox{E} \left[  \|{\bf A} {\mathbold \lambda}+ {\bf B}{ \bf Y}_{t-1}^{m}\|_{2}^r \right] =
 \mbox{E} \left[
 \Bigl[ \sum_{i=1}^p \left(  ({\bf A} {\mathbold \lambda})_i +  ({\bf B}{ \bf Y}_{t-1}^{m})_i  \right)^2 \Bigr] \right]^\mu
 := \mbox{E} \left(\sum_{i=1}^p C_i \right)^\mu,
\end{eqnarray*}
where $({\bf A} {\mathbold \lambda})_i $ and $({\bf B}{ \bf Y}_{t-1}^{m})_i $ are the $i$th components of the vectors
${\bf A} {\mathbold \lambda} $ and $ {\bf B}{ \bf Y}_{t-1}^{m}$, respectively. But
\begin{eqnarray*}
 \left(\sum_{i=1}^p C_i \right)^\mu = \sum_{i_{1}} \ldots \sum_{i_{p}} \frac{\mu!}{i_1! \ldots i_p!} C_1^{i_1} \ldots C_p^{i_p},
\end{eqnarray*}
where the sum extends over all indices $i_{j}, j=1,2,\ldots p$ such that $\sum_{j=1} ^p i_j = \mu$.
Successive use of the Cauchy-Schwartz inequality yields
\begin{eqnarray*}
 \mbox{E} \left(C_1^{i_1} \ldots C_p^{i_p}   \right) \leq   \mbox{E}^{1/2 l_{1}}(C_1^{2 i_{1} l_{1}}) \ldots \mbox{E}^{1/2 l_{p}}(C_p^{2 i_{p} l_{p}}),
\end{eqnarray*}
where  $ 1 \leq l_{p} \leq 2^{p-2}$, and
\begin{eqnarray*}
 \mbox{E}\left( C_k^{2i_k l_{k}} \right)  = \mbox{E} \left[ ({\bf A} {\mathbold \lambda})_k + ({\bf B}{ \bf Y}_{t-1}^{m})_k   \right]^{4 i_k l_{k}} =
 \mbox{E} \left[ \sum_{j=0}^{4 i_k l_{k}} {4 i_{k} l_{k} \choose j}  ({\bf A} {\mathbold \lambda})_k^j  ({\bf B}{ \bf Y}_{t-1}^{})_k^{4 i_k l_{k} -j}    \right].
\end{eqnarray*}
But using the reasoning on page 26 of \cite{Fokianoseral(2009)comp}, as ${\lambda}_k \rightarrow \infty $, $k=1, \ldots, p,$
\begin{eqnarray*}
 \mbox{E} \left[ ({\bf B}{ \bf Y}_{t-1}^{})_k^{4 i_k l_{k} -j} | {\mathbold \lambda}_{t-1}= {\mathbold \lambda} \right] \sim ({\bf{B}} {\mathbold \lambda})_k^{4 i_k l_{k} -j}.
\end{eqnarray*}
Hence
\begin{eqnarray*}
 \mbox{E}^{1/2 l_{k}} \left(  C_k^{2 i_k l_{k}} \right) \sim \left( ({\bf{A}} + {\bf{B}}) {\mathbold \lambda}   \right)_k^{2 i_k},
\end{eqnarray*}
and asymptotically $ \mbox{E} \left( C_1^{i_1} \ldots C_p^{i_p}  \right) \leq  \left( ({\bf{A}} + {\bf{B}}) {\mathbold \lambda}   \right)_1^{2 i_1}
\ldots \left( ({\bf{A}} + {\bf{B}}) {\mathbold \lambda}   \right)_p^{2 i_p}$. Therefore we obtain that
\begin{eqnarray*}
\mbox{E} \left( \sum_{i=1}^p C_i\right)^\mu &\leq& \sum_{i_{1}} \ldots \sum_{i_{p}} \frac{\mu!}{i_1! \ldots i_p!} \left[ \left( ({\bf{A}} + {\bf{B}}) {\mathbold \lambda}   \right)_1^{2} \right]^{i_1}
\ldots \left[ \left( ({\bf{A}} + {\bf{B}}) {\mathbold \lambda}   \right)_p^{2} \right]^{i_p} \\ &=& \left[\sum_{j=1}^p  \left( ({\bf{A}} + {\bf{B}}) {\mathbold \lambda}   \right)^{2}_j  \right]^\mu
= \left( \|| ({\bf A} + {\bf{B}}) {\mathbold \lambda} \||_{2}^{2} \right)^\mu \leq \left( \|| ({\bf A} + {\bf{B}})  \||_2^{2}  \|{\mathbold \lambda}\|_{2}^2    \right)^\mu
\end{eqnarray*}
which, using the Tweedie criterion as in \citet[Prop. 2.1]{Fokianosetal(2009)},  implies that $\|| {\bf A} + {\bf{B}}  \||_2 < 1$ is a sufficient condition, and the proposition thus holds.

\subsection{Proof of Lemma \ref{lemma approx linear}}

To prove the first item of the Lemma, note that
\begin{eqnarray*}
\|\mbox{E}({\mathbold{\lambda}}_{t}^{m}-{\mathbold{\lambda}}_{t})\|_{2} & = &
\|{\bf A} \mbox{E}( {\mathbold{\lambda}}_{t-1}^{m}-{\mathbold{\lambda}}_{t-1}) + {\bf B} \mbox{E}({ \bf Y}_{t-1}^{m}-{ \bf Y}_{t-1})
+ \mbox{E}({\mathbold{\epsilon}}_{t}^{m})\|_{2} \\
& = &
\| {\bf A}  \mbox{E}\left( {\mathbold{\lambda}}_{t-1}^{m}-{\mathbold{\lambda}}_{t-1}\right) +  {\bf B}
\left[ \mbox{E} \left[ \mbox{E} \left( ({ \bf Y}_{t-1}^{m} | {\cal{F}}_{t-1;m}^{ \bf{Y}, \mathbold{\lambda}} )\right) \right]    - \mbox{E} \left[ \mbox{E} \left( { \bf Y}_{t-1} | {\cal{F}}_{t-1}^{\bf{Y}, \mathbold{\lambda}}  \right)   \right] \right]+ \mbox{E}({\mathbold{\epsilon}}_{t}^{m})\|_{2}
\\& \leq &
\|| {\bf A} + {\bf B} \||_{2} \| \mbox{E}( {\mathbold{\lambda}}_{t-1}^{m}-{\mathbold{\lambda}}_{t-1})\|_{2} +
\|\mbox{E}({\mathbold{\epsilon}}_{t}^{m})\|_{2},
\end{eqnarray*}
where $ {\cal{F}}_{t-1}^{\bf{Y}, \mathbold{\lambda}} $ and $ {\cal{F}}_{t-1;m}^{\bf{Y}, \mathbold{\lambda}} $ are the $\sigma$-algebras generated by $\{ {\mathbold{\lambda}}_{s}, s \leq t  \ \}$ and
$\{{\mathbold{\lambda}}_{s}^m, s \leq t  \ \}$, respectively.
By recursion and the fact that $\|\mbox{E}({\mathbold{\epsilon}}_{t}^{m})\|_{2} \leq c_{m}$ which tends to zero as $m \rightarrow \infty$ we obtain the desired
result.
\noindent
To prove the second statement, note that as $m \rightarrow \infty$,
\begin{eqnarray*}
\mbox{E}\|({\mathbold{\lambda}}_{t}^{m}-{\mathbold{\lambda}}_{t})\|_{2}^{2} & \sim & \mbox{E}\|{\bf A} ({\mathbold{\lambda}}_{t-1}^{m}-{\mathbold{\lambda}}_{t-1})
+ {\bf B}  ({\bf Y}_{t-1}^{m}-{ \bf Y}_{t-1}) \|_{2}^2.
\end{eqnarray*}
Let $\Delta_{t-1} {\mathbold{\lambda}} =  {\mathbold{\lambda}}_{t-1}^{m}-{\mathbold{\lambda}}_{t-1}$ and $\Delta_{t-1} {\bf Y} =  {\bf Y}_{t-1}^{m}-{\bf Y}_{t-1}$ , then
\begin{eqnarray*}
\mbox{E}\|({\mathbold{\lambda}}_{t}^{m}-{\mathbold{\lambda}}_{t})\|_{2}^{2} & \sim &
\mbox{E} \left[ \Delta_{t-1} {\mathbold{\lambda}}^T  {\bf A}^T {\bf A} \Delta_{t-1} {\mathbold{\lambda}}  +
\Delta_{t-1} {\mathbold{\lambda}}^T  {\bf A}^T {\bf B} \Delta_{t-1} {\bf Y} +
\Delta_{t-1} {\bf Y}^T  {\bf B}^T {\bf A} \Delta_{t-1} {\mathbold{\lambda}}   +
 \Delta_{t-1} {\bf Y}^T  {\bf B}^T {\bf B} \Delta_{t-1} {\bf Y}  \right] \\ & =&
 \mbox{E} \left[ \Delta_{t-1} {\mathbold{\lambda}}^T  {\bf C} \Delta_{t-1} {\mathbold{\lambda}}  +
\Delta_{t-1} {\mathbold{\lambda}}^T  {\bf D} \Delta_{t-1} {\bf Y} +
\Delta_{t-1} {\bf Y}^T  {\bf D}^T \Delta_{t-1} {\mathbold{\lambda}}   +
 \Delta_{t-1} {\bf Y}^T  {\bf E} \Delta_{t-1} {\bf Y}  \right]  \\ & :=&
 \sum_{i=1}^p \sum_{j=1}^p \mbox{E} \left[ c_{ij} \Delta_{t-1} \lambda_i \Delta_{t-1} \lambda_j + d_{ij} \Delta_{t-1} \lambda_i \Delta_{t-1} Y_j
 + d_{ji} \Delta_{t-1} \lambda_i \Delta_{t-1} Y_j + e_{ij} \Delta_{t-1} Y_i \Delta_{t-1} Y_j \right],
\end{eqnarray*}
where ${\bf C}={\bf A}^{T} {\bf A}$, ${\bf D}={\bf A}^{T} {\bf B}$ and ${\bf E}={\bf B}^T {\bf B}$.
By using properties of conditional expectation as before, we obtain
\begin{eqnarray*}
 \mbox{E} \left[ d_{ij} \Delta_{t-1} \lambda_i \Delta_{t-1} Y_j
 + d_{ji} \Delta_{t-1} \lambda_i \Delta_{t-1} Y_j      \right] &=& \mbox{E} \left[d_{ij} \Delta_{t-1} \lambda_i \Delta_{t-1} \lambda_j
 + d_{ji} \Delta_{t-1} \lambda_i \Delta_{t-1} \lambda_j        \right]
\end{eqnarray*}
In addition, following the  proof in \citet[Lemma 2.1]{Fokianosetal(2009)}, and  using the above conditioning argument,
\begin{eqnarray*}
 \mbox{E} \left( \Delta_{t-1} Y_i^2    \right) = \mbox{E} \left( \Delta_{t-1} \lambda_i \right)^2 + 2 \delta_{i,m},
\end{eqnarray*}
where $\delta_{i,m} \rightarrow 0$, as $m \rightarrow \infty$.
For the cross-terms we have to condition on the copula structure, $ {\cal{F}}_{t-1;m}^{\bf{Y}, \mathbold{\lambda}} $, as well i.e.
\begin{eqnarray*}
 \mbox{E} \left( \Delta_{t-1} Y_i  \Delta_{t-1} Y_j   \right) & =& \mbox{E} \left[ \mbox{E} \left[  \Delta_{t-1} Y_i  \Delta_{t-1} Y_j  |
      {\cal{F}}_{t-1;m}^{\bf{Y}, \mathbold{\lambda}}, {\cal{F}}_{t-1}^{\bf{Y}, \mathbold{\lambda}}\right]  \right]
 = \mbox{E}\left( \Delta_{t-1} \lambda_i \Delta_{t-1} \lambda_j   \right).
\end{eqnarray*}
Collecting all previous results, we obtain
\begin{eqnarray*}
\mbox{E} \| \left( {\mathbold{\lambda}}_{t}^{m}-{\mathbold{\lambda}}_{t})\right) \|_{2}^{2} & = &
\mbox{E} \|  ({\bf A} + {\bf B}) ( {\mathbold{\lambda}}_{t-1}^{m}-{\mathbold{\lambda}}_{t-1})  \|_{2}^{2} + D_{m} \\
& \leq &
\|| {\bf A} + {\bf B} \||_{2}^{2}  \mbox{E}  \| ({\mathbold{\lambda}}_{t-1}^{m}-{\mathbold{\lambda}}_{t-1}) \|_{2}^{2} + D_{m},
\end{eqnarray*}
where $D_{m} \rightarrow 0$ as $m \rightarrow \infty$. The last two statements are proved using straightforward adaptation of the proof
of \citet[Lemma 2.1]{Fokianosetal(2009)}.

\subsection{Proof of Proposition  \ref{ergodicity linear perturbed wd}}

The proof is based on \citet[Thm. 3.1]{DoukhanandWintenberger(2008)} and parallels the proof given by \citet{Doukhanetal(2011)}.
In proving weak dependence, we define the $\mathbold{X}_{t}=(\mathbold{Y}^{T}_{t}, \mathbold{\lambda}^{T}_{t})^{T}$ and we employ
the norm $\| \mathbold{x} \|_{\epsilon}= \|\mathbold{y}\|_{1} + \epsilon \|\mathbold{\lambda}\|_{1}$, where $\epsilon$ is not necessarily small.
Then, the contraction property is verified by noting that    $\mathbold{X}_{t}=F( \mathbold{X}^{T}_{t-1}, \mathbold{N}^{T}_{t})$ where
$\mathbold{N}_{t}$ is an iid sequence of $p$-variate copula Poisson processes and choosing $\epsilon=\||\mathbold{A}\||_{1}/\|| \mathbold{B}\||_{1}$.
This proves that $\mbox{E}[\|\mathbold{Y}_{t}\|_{1}] < \infty$ and $\mbox{E}[\|\mathbold{\lambda}_{t}\|_{1}] < \infty$.

To show finiteness of moments we will be using induction and a different technique than the method used in \citet{Doukhanetal(2011)}.
More precisely, suppose that $\mbox{E}[\|\mathbold{Y}_{t}\|^{r-1}_{r-1}] < \infty$ and $\mbox{E}[\|\mathbold{\lambda}_{t}\|^{r-1}_{r-1}] < \infty$
for $r \in \mathbb{N}$ and $r > 1$. Then consider the $i$-th component of $\mathbold{Y}_{t}$.
But
\begin{eqnarray*}
\mbox{E} \Bigl[ Y^{r}_{i,t} \mid {\cal F}^{\mathbold{Y}, \mathbold{\lambda}}_{t-1} \Bigr] & \leq &
\mbox{E}\Bigl[(Y_{i, t})_{r} \mid {\cal F}^{\mathbold{Y}, \mathbold{\lambda}}_{t-1} \Bigr]
+\sum_{k=1}^{r-1} | \delta_{ik}(r)| \mbox{E}\Bigl[Y^{k}_{i, t}\mid {\cal F}^{\mathbold{Y}, \mathbold{\lambda}}_{t-1} \Bigr] \\
& = &
\lambda^{r}_{i,t}+\sum_{k=1}^{r-1} | \delta_{ik}(r)| \mbox{E}\Bigl[Y^{k}_{i, t} \mid {\cal F}^{\mathbold{Y}, \mathbold{\lambda}}_{t-1} \Bigr],
\end{eqnarray*}
where $(x)_{r}=x(x-1)....(x-r+1)$, $\{ \delta_{jk}(r), ~k=1,2,\ldots, r-1\}$ are some constants and the first line follows from
properties of $(x)_{r}$ while the second line follows form properties of the Poisson distribution. By taking expectations
and using the $c_{r}$-inequality, we obtain that
\begin{eqnarray*}
\mbox{E}^{1/r} \Bigl[ Y^{r}_{i,t} \Bigr] & \leq &
\mbox{E}^{1/r} \Bigl[ \lambda^{r}_{i,t} \Bigr] + \sum_{k=1}^{r-1} | \delta_{ik}(r)|^{1/r}  \mu_{i},
\end{eqnarray*}
where $\mu_{i}=\max_{k <r} \mbox{E}\Bigl[Y^{k}_{i, t} \mid {\cal F}^{\mathbold{Y}, \mathbold{\lambda}}_{t-1} \Bigr]$,
which exists
by the induction hypothesis. But
\begin{eqnarray*}
\mbox{E} (\lambda_{i,t}^{r})  \leq \mbox{E}( Y_{i,t}^{r}),
\end{eqnarray*}
because of the properties of the linear model. Therefore, we obtain that (because of \eqref{linear model in terms of Poisson})
\begin{eqnarray*}
\mbox{E}^{1/r} [Y_{i,t}^{r} ] \leq d_{i} + \sum_{j=1}^{p} a_{ij} \mbox{E}^{1/r} [Y_{i,t}^{r} ]
+ \sum_{j=1}^{p} b_{ij} \mbox{E}^{1/r} [Y_{i,t}^{r} ] +  \sum_{k=1}^{r-1}
|\delta_{ik}(r)|^{1/r} \mu_{i},
\end{eqnarray*}
and by summing up, using the definition of $\|| . \||_{1}$ and its properties, we obtain that
\begin{eqnarray*}
\sum_{i=1}^{p}\mbox{E}^{1/r} [Y_{i,t}^{r} ] \leq  \sum_{i=1}^{p} d_{i} + (\||A\||_{1}+\||B\||_{1}) \sum_{i=1}^{p} \mbox{E}^{1/r} [Y_{i,t}^{r} ] +  \sum_{i=1}^{p}\sum_{k=1}^{r-1} | \delta_{ik}(r)|^{1/r} \mu_{i}.
\end{eqnarray*}

\subsection{Proof of Lemma \ref{lemma approx loglinear}}

\noindent
We will  prove the second and fourth  conclusion as the other results follow from \citet{FokianosandTjostheim(2011)}
and the proof of Lemma \ref{lemma approx linear}. But
to prove the second statement, note that
\begin{eqnarray*}
\mbox{E}\|({\mathbold{\nu}}_{t}^{m}-{\mathbold{\nu}}_{t})\|_{2}^{2} & = &
\mbox{E}\|{\bf A} \mbox{E}( {\mathbold{\nu}}_{t-1}^{m}-{\mathbold{\nu}}_{t-1}) + {\bf B} \mbox{E}(\log({ \bf Y}_{t-1}^{m}+\mathbold{1}_{p})-\log({ \bf Y}_{t-1}+\mathbold{1}_{p}))
+ \mbox{E}({\mathbold{\epsilon}}_{t}^{m})\|_{2}^{2} \\
& \leq &
\|| {\bf A} \||_{2}^{2} \mbox{E}\| {\mathbold{\nu}}_{t-1}^{m}-{\mathbold{\nu}}_{t-1}\|_{2}^{2}+
\|| {\bf B} \||_{2}^{2} \mbox{E}\|\log({ \bf Y}_{t-1}^{m}+\mathbold{1}_{p})-\log({ \bf Y}_{t-1}+\mathbold{1}_{p})\|_{2}^{2} \\
& + &
2 \|| {\bf A} \||_{2} \|| {\bf B} \||_{2}
\sqrt{ \mbox{E}\| {\mathbold{\nu}}_{t-1}^{m}-{\mathbold{\nu}}_{t-1}\|_{2}^{2} \mbox{E}\|\log({ \bf Y}_{t-1}^{m}+\mathbold{1}_{p})-\log({ \bf Y}_{t-1}+\mathbold{1}_{p})|_{2}^{2}}+ \kappa c_{m}^{2},
\end{eqnarray*}
where $\kappa > 0$. Consider now the $\mbox{E}(\log(Y_{j,t-1}^{m}+1)-\log(Y_{j, t-1}))^{2}$, $j=1,2,\ldots,p$. Then, following the proof of
\citet[Lemma 2.1]{FokianosandTjostheim(2011)} and assuming without loss of generality
that $\lambda_{j, t-1}^{m} \geq \lambda_{j, t-1}$ we obtain that $((Y_{j,t-1}^{m}+1)/(Y_{j,t-1}+1) \geq 1$. Therefore
by using Jensen's inequality (by employing the function $(\log x)^{2}$) we obtain that
\begin{eqnarray*}
\mbox{E} \left[ \log \frac{Y_{j,t-1}^{m}+1}{Y_{j, t-1}+1} \right]^{2} \leq
\left[ \log \mbox{E} \left(\frac{Y_{j,t-1}^{m}+1}{Y_{j, t-1}+1} \right)^{2} \right].
\end{eqnarray*}
But according to \citet[p. 576]{FokianosandTjostheim(2011)} the right hand side of the above inequality is bounded
by $\mbox{E}( \nu_{j, t-1}^{m}- \nu_{j;t-1})^{2}$ for $j=1,2,\ldots,p$. Hence, the conclusion of the Lemma follows again
by the same arguments used in the proof of Lemma \ref{lemma approx linear}.

\noindent
To prove the fourth result, we follow \citet[pp. 576-577]{FokianosandTjostheim(2011)}.
Consider the test function $V({\bf x})= \exp( r \|{\bf x}\|_{2})$  for $r \in \mathbb{N}$.  Set  $b=r \|| \mathbold{B} \||_{2}$.
Then
\begin{eqnarray*}
\mbox{E}[ \exp( r \| \mathbold{\nu}^{m}_{t} \|_{2}) \mid \mathbold{\nu}^{m}_{t-1}=\mathbold{\nu}]
& \leq &
\exp( r( \|\mathbold{d}\|_{2} + \||\mathbold{A}\||_{2} \| \mathbold{\nu} \|_{2}) )
\mbox{E}[ \exp(r \||\mathbold{B}\||_{2} \| \log(\mathbold{Y}^{m}_{t-1}+ \mathbold{1}_{p}) \|_{2}  ) \mid \mathbold{\nu}^{m}_{t-1}=\mathbold{\nu}].
\end{eqnarray*}
However
\begin{eqnarray*}
\mbox{E} \Bigl[ \exp \Bigl(b \| \log(\mathbold{Y}^{m}_{t-1}+ \mathbold{1}_{p}) \|_{2} \Bigr) |  \mathbold{\nu}^{m}_{t-1}= \mathbold{\nu} \Bigr] & = &
\mbox{E} \left\{ \exp \Bigl[ b \Bigl( \sum_{i=1}^{p} \log^{2}(Y^{m}_{i, t-1}+1)\Bigr)^{1/2} \Bigr]  | \mathbold{\nu}^{m}_{t-1}= \mathbold{\nu} \right\}\nonumber \\
& = &
\mbox{E} \left\{ \exp \Bigl[ b  \Bigl( \sum_{i=1}^{p} \Bigl(  \nu_{i}+
\Bigl( \frac{\log(Y^{m}_{i, t-1}+1)}{\exp(\nu_{i})} \Bigr) \Bigr)^{2} \Bigr)^{1/2}  \bigl]|
\mathbold{\nu}^{m}_{t-1}= \mathbold{\nu} \right\}.
\end{eqnarray*}
But
\begin{eqnarray}
\mbox{Var} \Bigl[  \frac{Y^{m}_{t}+1}{\exp(\nu_{i})}  \mid  \mathbold{\nu}^{m}_{t-1}= \mathbold{\nu} \Bigr]=
\exp(- \nu_{i}) \rightarrow 0,
\label{varapprox}
\end{eqnarray}
provided that $\nu_{i} \rightarrow \infty$ for all $i=1,2,\ldots,p$. Therefore we have that
\begin{eqnarray*}
\mbox{Var} \Bigl[ \log \Bigr( \frac{Y^{m}_{t}+1}{\exp(\nu_{i})} \Bigl)  \mid  \mathbold{\nu}^{m}_{t-1}= \mathbold{\nu} \Bigr]
\rightarrow 0,
\end{eqnarray*}
by the delta-method for moments and provided that $\nu_{i} \rightarrow \infty$ for all $i=1,2,\ldots,p$. Using now the multivariate delta-method
and Cauchy-Schwartz inequality to the function $g(x_{1}, \ldots, x_{p}) = \exp( b (\sum_{i}^{p}(\nu_{i}+x_{i})^{2})^{1/2})$ (with some abuse
of notation), we obtain that
\begin{eqnarray*}
\mbox{Var} \left\{ \exp \Bigl[ b  \Bigl( \sum_{i=1}^{p} \Bigl(  \nu_{i}+
\Bigl( \frac{\log(Y^{m}_{i, t-1}+1)}{\exp(\nu_{i})} \Bigr) \Bigr)^{2} \Bigr)^{1/2}  \bigl]|
\mathbold{\nu}^{m}_{t-1}= \mathbold{\nu} \right\} \rightarrow 0.
\end{eqnarray*}
However
\begin{eqnarray}
\mbox{E} \Bigl[  \frac{Y^{m}_{t}+1}{\exp(\nu_{i})}  \mid  \mathbold{\nu}^{m}_{t-1}= \mathbold{\nu} \Bigr] \sim 1
\label{expectapprox}
\end{eqnarray}
provided that $\nu_{i} \rightarrow \infty$ for all $i=1,2,\ldots,p$. Therefore, asymptotically, we obtain  that
\begin{eqnarray*}
\mbox{E} \Bigl[ \exp \Bigl(b \| \log(\mathbold{Y}^{m}_{t-1}+ \mathbold{1}_{p}) \|_{2} \Bigr) |  \mathbold{\nu}^{m}_{t-1}= \mathbold{\nu} \Bigr]
\sim \exp(b \| \mathbold{\nu} \|_{2}).
\end{eqnarray*}
To complete the proof, we note that the above calculations show that
\begin{eqnarray*}
\mbox{E} \Bigl[ \exp( r \| \mathbold{\nu}^{m}_{t})\|_{2})  |  \mathbold{\nu}^{m}_{t-1}= \mathbold{\nu}]  \leq
\exp( r(\|| \mathbold{A}_{2} \||_{2} + \|| \mathbold{B}_{2} \||_{2}-1) \| \mathbold{\nu} \|_{2} ) \exp( r\| \mathbold{\nu} \|_{2}).
\end{eqnarray*}
Therefore, the conclusion follows as in \citet[pp. 576-577]{FokianosandTjostheim(2011)}.

\subsection{Proof of Proposition \ref{ergodicity loglinear perturbed wd}}

For the log-linear model we prove weak dependence by the following method.  Set
$Y_{j,t} = N_{j, t} ( \exp( \nu_{j, t}))$, $j=1,2,\ldots,p$. Then setting $Z_{j,t} =\log(1+Y_{j,t})$ we have for $\mathbold{X}_{t}=(\mathbold{Z}_{t}, \mathbold{\nu}_{t})$
with $\mathbold{Z}_{t}= (Z_{j,t}, j=1,2,\ldots,p)$ and  $N_{t}= (N_{j,t}, j=1,2, \ldots,d)$
that
\begin{eqnarray*}
\mathbold{X}_{t} = (\mathbold{Z}_{t}, \mathbold{\nu}_{t}) = F(\mathbold{X}^{T}_{t-1}, \mathbold{N}^{T}_{t}), 
\end{eqnarray*}
where $\mathbold{N}_{t}= (N_{j,t}, j=1,2, \ldots,p)$ iid copula $p$-variate Poisson processes.
Then using  again the same arguments as in \citet{Doukhanetal(2011)}  we obtain (with the same norm) that
\begin{eqnarray*}
\mbox{E} \left[ \| F( \mathbold{x}, \mathbold{N})- F( \mathbold{x}^{\star}, \mathbold{N}) \|_{\epsilon} \right]  &  \leq &  \sum_{j=1}^{p}
\|\Bigl( \mathbold{A}(\mathbold{\nu}-\mathbold{\nu}^{\star})\Bigr)_{j}\|_{1} +
\sum_{j=1}^{p} \|\Bigl( \mathbold{B} (\mathbold{\zeta}-\mathbold{\zeta}^{\star})\Bigr)_{j} \|_{1}  \\ \nonumber
& + &   \epsilon \Bigl( \| \mathbold{A} ( \nu-\nu^{\star})\|_{1} + \| \mathbold{B} (\mathbold{\zeta}-\mathbold{\zeta}^{\star})\|_{1} \Bigr)
\nonumber \\
& \leq &  (1+\epsilon) \Bigl( \||\mathbold{A} \||_{1} \mathbold{\nu}-\mathbold{\nu}^{\star}\|_{1} + \||\mathbold{B}\||_{1} \| \mathbold{\zeta}-\mathbold{\zeta}^{\star}\|_{1} \Bigr),
\end{eqnarray*}
where the first inequality  follows from  \citet[pp.575--576]{FokianosandTjostheim(2011)}. The results now follow
as in  \citet{Doukhanetal(2011)}.
Now we show existence of moments for the log-linear model. Suppose that $r \in \mathbb{N}$. Then
\begin{eqnarray*}
\mbox{E}[ \exp( r \| \mathbold{\nu}_{t} \|_{1}) \mid \mathbold{\nu}_{t-1}=\mathbold{\nu}]
& \leq &
\exp( r( \|\mathbold{d}\|_{1} + \||\mathbold{A}\||_{1} \| \mathbold{\nu} \|_{1}) )
\mbox{E}[ \exp(r \||\mathbold{B}\||_{1} \|\mathbold{Z}_{t-1} \|_{1}  ) \mid \mathbold{\nu}_{t-1}=\mathbold{\nu}] \nonumber
\end{eqnarray*}
With $b=r \|| \mathbold{B} \||_{1}$, for   the second factor of the right hand side we obtain that
\begin{eqnarray*}
\mbox{E} \Bigl[ \exp\Bigl( b \|\mathbold{Z}_{t-1} \|_{1} \Bigr) \mid \mathbold{\nu}_{t-1}=\mathbold{\nu} \Bigr]  =
\exp(b \| \mathbold{\nu} \|_{1})
\mbox{E} \Bigl[
\prod_{i=1}^{p}  \Bigl( \frac{Y_{i,t-1}+1}{ \exp(\nu_{i})} \Bigr)^{b} \mid \mathbold{\nu}_{t-1}=\mathbold{\nu} \Bigr]
\end{eqnarray*}
But from the proof of Lemma \ref{lemma approx loglinear} (see eq. \eqref{varapprox}) and using similar arguments
\begin{eqnarray*}
\mbox{Var} \Bigl[  \prod_{i=1}^{p}  \Bigl( \frac{Y_{i,t-1}+1}{ \exp(\nu_{i})} \Bigr)^{b} \mid \mathbold{\nu}_{t-1}=\mathbold{\nu} \Bigr]
\rightarrow 0,
\end{eqnarray*}
provided that $\nu_{i} \rightarrow \infty$, for all $i=1,2\ldots, p$. In addition, because of  \eqref{expectapprox} and the multivariate
delta-method of moments
\begin{eqnarray*}
\mbox{E} \Bigl[  \prod_{i=1}^{p}  \Bigl( \frac{Y_{i,t-1}+1}{ \exp(\nu_{i})} \Bigr)^{b} \mid \mathbold{\nu}_{t-1}=\mathbold{\nu} \Bigr]
\rightarrow 1 ,
\end{eqnarray*}
provided that $\nu_{i} \rightarrow \infty$, for all $i=1,2\ldots, p$. The above two displays show that
\begin{eqnarray*}
\mbox{E} \Bigl[ \exp\Bigl( b \|\mathbold{Z}_{t-1} \|_{1} \Bigr) \mid \mathbold{\nu}_{t-1}=\mathbold{\nu} \Bigr]  \sim
\exp(b \| \mathbold{\nu} \|_{1}),
\end{eqnarray*}
as required.

%

\subsection{Proof of Lemma \ref{Lemma1qmle}}

In what follows we drop notation that depends on $\mathbold{\theta}$ because all quantities are evaluated at the true parameter $\mathbold{\theta}_{0}$.
The notation $C$ refers to a generic constant. Initially, we show that
\begin{eqnarray}
\Bigl\| \Bigl| \frac{\partial \mathbold{\lambda}_{t}^{m}}{\partial \mathbold{d}^{T}} - \frac{\partial \mathbold{\lambda}_{t}}{\partial \mathbold{d}^{T}}
\Bigr\| \Bigr|_{2}
< \gamma_{m}, ~ a.s,
\label{difference der wrt to d}
\end{eqnarray}
for some positive sequence $\gamma_{m} \rightarrow 0$, as $m \rightarrow \infty$. Using the first equation of \eqref{recursions linear}
we obtain that
\begin{eqnarray*}
\Bigl\| \Bigl|  \frac{\partial \mathbold{\lambda}_{t}^{m}}{\partial \mathbold{d}^{T}} - \frac{\partial \mathbold{\lambda}_{t}}{\partial \mathbold{d}^{T}}
\Bigr\| \Bigr|_{2}
& \leq  & \||\mathbold{A}\||_{2}
\Bigl\| \Bigl|
\frac{\partial \mathbold{\lambda}_{t-1}^{m}}{\partial \mathbold{d}^{T}} - \frac{\partial \mathbold{\lambda}_{t-1}}{\partial \mathbold{d}^{T}}
\Bigr\| \Bigr|_{2}
\end{eqnarray*}
and therefore, by repeated substitution, \eqref{difference der wrt to d} follows since $ \||\mathbold{A}\||_{2} <1$ and the results of
Lemma \ref{lemma approx linear}.
Similarly,
\begin{eqnarray}
\Bigl\| \Bigl|
\frac{\partial \mathbold{\lambda}_{t}^{m}}{\partial \vect^{T}(\mathbold{A})} - \frac{\partial \mathbold{\lambda}_{t}}{\partial \vect^{T}(\mathbold{A})}
\Bigr\| \Bigr|_{2}  \leq \gamma_{m}, ~ a.s.
\label{difference der wrt to A}
\end{eqnarray}
Indeed, using the second equation of \eqref{recursions linear}, we obtain that
\begin{eqnarray*}
\Bigl\| \Bigl|
\frac{\partial \mathbold{\lambda}_{t}^{m}}{\partial \vect^{T}(\mathbold{A})} - \frac{\partial \mathbold{\lambda}_{t}}{\partial \vect^{T}(\mathbold{A})}
\Bigr\| \Bigr|_{2}
& \leq & \sqrt{p} \| \mathbold{\lambda}_{t-1}^{m}-\mathbold{\lambda}_{t-1}\|_{2} + \||\mathbold{A}\||_{2}
\Bigl\| \Bigl|
\frac{\partial \mathbold{\lambda}_{t-1}^{m}}{\partial \vect^{T}(\mathbold{A})} - \frac{\partial \mathbold{\lambda}_{t-1}}{\partial \vect^{T}(\mathbold{A})}
\Bigr\| \Bigr|_{2},
\end{eqnarray*}
where the first bound comes from the fact that in terms of the Frobenious matrix norm  $\|| \mathbold{I}_{p} \||_{F}= \sqrt{p}$.
Therefore, by Lemma \ref{lemma approx linear} we obtain the desired result. Finally, it can be shown quite analogously (by using
again Lemma \eqref{lemma approx linear}) that
\begin{eqnarray}
\Bigl\| \Bigl|
\frac{\partial \mathbold{\lambda}_{t}^{m}}{\partial \vect^{T}(\mathbold{B})} - \frac{\partial \mathbold{\lambda}_{t}}{\partial \vect^{T}(\mathbold{B})}
\Bigr\| \Bigr|_{2}
\leq \gamma_{m}, ~ a.s.
\label{difference der wrt to B}
\end{eqnarray}

To prove the lemma, we consider the $d \times d$ matrix difference
\begin{eqnarray}
\Bigl\| \Bigl| s_{t}^{m}(s_t^m)^T - s_{t} s_{t}^{T} \Bigr\| \Bigr|_{2} & = & \Bigl\| \Bigl| (s_{t}^{m}-s_{t})(s_t^m)^T+ s_{t}(s_{t}^{m}-s_{t})^{T} \Bigr\| \Bigr|_{2} \nonumber \\
                                                                       & \leq & \| s_{t}^{m}-s_{t} \|_{2}  \| (s_t^m)^T\|_{2} +\|s_{t}\|_{2} \|(s_{t}^{m}-s_{t})^{T} \|_{2}.
\label{decompositionmain}
\end{eqnarray}
But
\begin{eqnarray}
s_{t}^{m}- s_{t} & = &
\Bigl[ \Bigl( \frac{\partial \mathbold{\lambda}_{t}^{m}}{\partial \mathbold{\theta}^{T}} \Bigr)^{T}
      - \Bigl(\frac{\partial \mathbold{\lambda}_{t}}{\partial \mathbold{\theta}^{T}} \Bigr)^{T} \Bigr]
\Bigl( \mathbold{D}_{t}^{m} \Bigr)^{-1}
\Bigl( \mathbold{Y}_{t}^{m}- \mathbold{\lambda}_{t}^{m}  \Bigr)   \nonumber \\
& + &
\Bigl( \frac{\partial \mathbold{\lambda}_{t}}{\partial \mathbold{\theta}^{T}} \Bigr)^{T}
\Bigl[ \Bigl( \mathbold{D}_{t}^{m} \Bigr)^{-1} -\Bigl( \mathbold{D}_{t} \Bigr)^{-1} \Bigr]
\Bigl( \mathbold{Y}_{t}^{m}- \mathbold{\lambda}_{t}^{m} \Bigr) \nonumber  \\
& + &
\Bigl( \frac{\partial \mathbold{\lambda}_{t}}{\partial \mathbold{\theta}^{T}} \Bigr)^{T}
\mathbold{D}_{t}^{-1}
\Bigl[ \Bigl( \mathbold{Y}_{t}^{m}- \mathbold{\lambda}_{t}^{m} \Bigr)-\Bigl( \mathbold{Y}_{t}- \mathbold{\lambda}_{t}) \Bigr)   \nonumber \\
& = & (I)+ (II) +(III), \label{decomposition1Lemma1}
\end{eqnarray}
with obvious notation. Then we obtain for the first term $(I)$ of \eqref{decomposition1Lemma1}
\begin{eqnarray}
\| \Bigl[ \Bigl( \frac{\partial \mathbold{\lambda}_{t}^{m}}{\partial \mathbold{\theta}^{T}} \Bigr)^{T}
      - \Bigl(\frac{\partial \mathbold{\lambda}_{t}}{\partial \mathbold{\theta}^{T}} \Bigr)^{T} \Bigr]
\Bigl( \mathbold{D}_{t}^{m} \Bigr)^{-1}
\Bigl( \mathbold{Y}_{t}^{m}- \mathbold{\lambda}_{t}^{m} \Bigr)   \|_{2} & \leq \displaystyle{
 \Bigl\| \Bigl|   \frac{\partial \mathbold{\lambda}_{t}^{m}}{\partial \mathbold{\theta}^{T}}-\frac{\partial \mathbold{\lambda}_{t}}{\partial \mathbold{\theta}^{T}} \Bigr\| \Bigr|_{2}
\Bigl\|\Bigl|  \Bigl( \mathbold{D}_{t}^{m}  \Bigr)^{-1} \Bigr\| \Bigr|_{2}
\| \Bigl( \mathbold{Y}_{t}^{m}- \mathbold{\lambda}_{t}^{m} \Bigr)\|_{2}}. \nonumber \\
\label{decomposition1step1Lemma}
\end{eqnarray}
We deal with the first factor. Recall that $\||. \||_{F}$ stands for the Frobenious norm of a matrix. Then
\begin{eqnarray*}
 \Bigl\| \Bigl|   \frac{\partial \mathbold{\lambda}_{t}^{m}}{\partial \mathbold{\theta}^{T}}-\frac{\partial \mathbold{\lambda}_{t}}{\partial\mathbold{\theta}^{T}} \Bigr\| \Bigr|_{2}^{2}
& \leq &
\Bigl\| \Bigl|   \frac{\partial \mathbold{\lambda}_{t}^{m}}{\partial \mathbold{\theta}^{T}}-\frac{\partial \mathbold{\lambda}_{t}}{\partial \mathbold{\theta}^{T}} \Bigr\| \Bigr|_{F}^{2} \\
& = &
 \Bigl\| \Bigl|   \frac{\partial \mathbold{\lambda}_{t}^{m}}{\partial \mathbold{d}^{T}}-\frac{\partial \mathbold{\lambda}_{t}}{\partial \mathbold{d}^{T}} \Bigr\| \Bigr|_{F}^{2}
+
\Bigl\| \Bigl|
\frac{\partial \mathbold{\lambda}_{t}^{m}}{\partial \vect^{T}(\mathbold{A})} - \frac{\partial \mathbold{\lambda}_{t}}{\partial \vect^{T}(\mathbold{A})}^{2}
\Bigr\| \Bigr|_{F}
+
\Bigl\| \Bigl|
\frac{\partial \mathbold{\lambda}_{t}^{m}}{\partial \vect^{T}(\mathbold{B})} - \frac{\partial \mathbold{\lambda}_{t}}{\partial \vect^{T}(\mathbold{B})}
\Bigr\| \Bigr|_{F}^{2}  \\
&  \leq &
p \Bigl\| \Bigl|   \frac{\partial \mathbold{\lambda}_{t}^{m}}{\partial \mathbold{d}^{T}}-\frac{\partial \mathbold{\lambda}_{t}}{\partial \mathbold{d}^{T}} \Bigr\| \Bigr|_{2}^{2}
+
p^{2} \Bigl\| \Bigl|
\frac{\partial \mathbold{\lambda}_{t}^{m}}{\partial \vect^{T}(\mathbold{A})} - \frac{\partial \mathbold{\lambda}_{t}}{\partial \vect^{T}(\mathbold{A})}
\Bigr\| \Bigr|_{2}^{2}
+
p^{2} \Bigl\| \Bigl|
\frac{\partial \mathbold{\lambda}_{t}^{m}}{\partial \vect^{T}(\mathbold{B})} - \frac{\partial \mathbold{\lambda}_{t}}{\partial \vect^{T}(\mathbold{B})}
\Bigr\| \Bigr|_{2}^{2},
\end{eqnarray*}
where the first and third inequality hold because of result 4.67(a) of \citet{Seber(2008)} and the second inequality is a consequence of the definition
of Frobenious norm.  Then we need to  show that
\begin{eqnarray}
\mbox{E} \Bigl[
\Bigl\| \Bigl|   \frac{\partial \mathbold{\lambda}_{t}^{m}}{\partial \mathbold{d}^{T}}-\frac{\partial \mathbold{\lambda}_{t}}{\partial \mathbold{d}^{T}} \Bigr\| \Bigr|_{2}^{2}
\Bigr], ~~
\mbox{E}  \Bigl[
\Bigl\| \Bigl|
\frac{\partial \mathbold{\lambda}_{t}^{m}}{\partial \vect^{T}(\mathbold{A})} - \frac{\partial \mathbold{\lambda}_{t}}{\partial \vect^{T}(\mathbold{A})}
\Bigr\| \Bigr|_{2}^{2}
\Bigr], ~~
\mbox{E}  \Bigl[
\Bigl\| \Bigl|
\frac{\partial \mathbold{\lambda}_{t}^{m}}{\partial \vect^{T}(\mathbold{B})} - \frac{\partial \mathbold{\lambda}_{t}}{\partial \vect^{T}(\mathbold{B})}
\Bigr\| \Bigr|_{2}^{2}
\Bigr]    \leq   \gamma_{m},
\label{decomposition1step1aLemma}
\end{eqnarray}
with  $\gamma_{m} \rightarrow 0$.
We deal with the middle term only; similar arguments can be used for the other two terms.
Squaring  the expression after \eqref{difference der wrt to A} and taking  expectations we obtain that
\begin{eqnarray*}
\mbox{E} \Bigl[
\Bigl\| \Bigl|
\frac{\partial \mathbold{\lambda}_{t}^{m}}{\partial \vect^{T}(\mathbold{A})} - \frac{\partial \mathbold{\lambda}_{t}}{\partial \vect^{T}(\mathbold{A})}
\Bigr\| \Bigr|_{2}^{2}
\Bigr]
& \leq &
p \mbox{E} \Bigl[
\| \mathbold{\lambda}_{t-1}^{m}-\mathbold{\lambda}_{t-1}\|_{2}^{2} \Bigr]
+
 \||\mathbold{A}\||_{2}^{2}
\mbox{E} \Bigl[\Bigl\| \Bigl|
\frac{\partial \mathbold{\lambda}_{t-1}^{m}}{\partial \vect^{T}(\mathbold{A})} - \frac{\partial \mathbold{\lambda}_{t-1}}{\partial \vect^{T}(\mathbold{A})}
\Bigr\| \Bigr|_{2}^{2} \Bigr] \\
& + & 2 \sqrt{p} \||\mathbold{A}\||_{2} \mbox{E} \Bigl[
\| \mathbold{\lambda}_{t-1}^{m}-\mathbold{\lambda}_{t-1}\|_{2}
\Bigl\| \Bigl|
\frac{\partial \mathbold{\lambda}_{t-1}^{m}}{\partial \vect^{T}(\mathbold{A})} - \frac{\partial \mathbold{\lambda}_{t-1}}{\partial \vect^{T}(\mathbold{A})}
\Bigr\| \Bigr|_{2} \Bigr] \\
& \leq &  \delta_{2;m} + \||\mathbold{A}\||_{2}^{2}
\mbox{E} \Bigl[\Bigl\| \Bigl|
\frac{\partial \mathbold{\lambda}_{t-1}^{m}}{\partial \vect^{T}(\mathbold{A})} - \frac{\partial \mathbold{\lambda}_{t-1}}{\partial \vect^{T}(\mathbold{A})}
\Bigr\| \Bigr|_{2}^{2} \Bigr]  + 2  C \sqrt{p} \||\mathbold{A}\||_{2} \sqrt{ \delta_{2;m}} \leq \gamma_{m},
\end{eqnarray*}
where $\gamma_{m}$ can become arbitrarily small. This follows from Proposition \ref{ergodicity linear perturbed}, \eqref{difference der wrt to A} and the fact that
$\||\mathbold{A}\||_{2} < 1$.
For the second term of \eqref{decomposition1step1Lemma} we note that
\begin{eqnarray}
\Bigl\| \Bigl| \Bigl( \mathbold{D}_{t}^{m} \Bigr)^{-1} \Bigr\| \Bigr|_{2} \leq \sqrt{p  \max_{1 \leq i \leq p} \frac{1}{d^{2}_{i}}} \leq C,
\label{decomposition1step2Lemma}
\end{eqnarray}
where $d_{i}$ is the $i$'th component of $\mathbold{d}$. In addition
\begin{eqnarray}
\mbox{E}\Bigl[ \|\Bigl( \mathbold{Y}_{t}^{m}- \mathbold{\lambda}_{t}^{m} \Bigr)\|_{2}^{2} \Bigr]
=
\sum_{i=1}^{p} \mbox{E}[ \lambda_{i,t}^{m}] <  C
\label{decomposition1step3Lemma}
\end{eqnarray}
by  Proposition \ref{ergodicity linear perturbed} and using a conditioning argument. Collecting \eqref{decomposition1step1aLemma},
\eqref{decomposition1step2Lemma} and \eqref{decomposition1step3Lemma} an application of Cauchy-Schwartz inequality shows that the
$$
\mbox{E} \Bigl[
 \Bigl\| \Bigl|   \frac{\partial \mathbold{\lambda}_{t}^{m}}{\partial \mathbold{\theta}^{T}}-\frac{\partial \mathbold{\lambda}_{t}}{\partial \mathbold{\theta}^{T}} \Bigr\| \Bigr|_{2}
\Bigl\|\Bigl|  \Bigl( \mathbold{D}_{t}^{m}  \Bigr)^{-1} \Bigr\| \Bigr|_{2}
\| \Bigl( \mathbold{Y}_{t}^{m}- \mathbold{\lambda}_{t}^{m} \Bigr)\|_{2}
\Bigr] \rightarrow 0,
$$
as $m \rightarrow \infty$.
Now we look at the second summand $(II)$ of \eqref{decomposition1Lemma1}. First of all, we note that
$
\mbox{E} \Bigl\| \Bigl|  {\partial \mathbold{\lambda}_{t}}/{\partial \mathbold{\theta}^{T}} \Bigr\| \Bigr|_{2}^{4} < C.
$
This is proved by using the same decomposition of the norm  as the sum of norms of the matrix of derivatives with respect to $\mathbold{d}$, $ \vect(\mathbold{A})$ and $ \vect(\mathbold{B})$.
Then using \eqref{recursions linear}, the fact that $\Bigl\| \Bigl| \mathbold{A} \Bigr\| \Bigr|_{2} < 1$ and  the compactness of the parameter space, the result follows.
In addition, for some finite constants  $(c_{ij})$, we obtain  that
\begin{eqnarray*}
\mbox{E} \Bigl[ \| \mathbold{Y}_{t}^{m}- \mathbold{\lambda}_{t}^{m}\|_{2}^{4} \Bigr] & = &
\mbox{E}  \Bigl[ \Bigl( \sum_{i=1}^{p} (Y_{i,t}^{m} - \lambda_{i,t}^{m})^{2} \Bigr)^{2} \Bigr]  \\
& = &
\mbox{E}
\Bigl[  \sum_{i=1}^{p} ( (Y_{i,t}^{m} - \lambda_{i,t}^{m})^{4} + \sum_{i \neq j}  (Y_{i,t}^{m} - \lambda_{i,t}^{m})^{2}  (Y_{j,t}^{m} - \lambda_{j,t}^{m})^{2} \Bigr] \\
& \leq &
\sum_{i=1}^{p} \mbox{E}[ (\lambda_{i,t}^{m})^{4}] + \sum_{i=1}^{p} \sum_{j=1}^{3}  c_{ij} \mbox{E}[ (\lambda_{i,t}^{m})^{j}] < C,
\end{eqnarray*}
because of Proposition \ref{ergodicity linear perturbed} and  from the same arguments given in the proof of Proposition \ref{ergodicity linear perturbed wd}.
Now we have that
\begin{eqnarray*}
\Bigl\| \Bigl| \Bigl( \mathbold{D}_{t}^{m} \Bigr)^{-1} -\Bigl( \mathbold{D}_{t} \Bigr)^{-1} \Bigl\| \Bigr|_{2}^{2} & \leq C \| \mathbold{\lambda}^{m}_{t}- \mathbold{\lambda}_{t}\|_{2}^{2}
\end{eqnarray*}
and therefore its expected value tends to zero by Lemma \ref{lemma approx linear}. Collecting all these results we have that the expected value of $(II)$ in
\eqref{decomposition1Lemma1} tends to zero.
Finally, the expected value of  term $(III)$ in \eqref{decomposition1Lemma1} tends to zero, as $m \rightarrow \infty$ by combining the
above results and using Cauchy-Schwartz inequality and Lemma  \ref{lemma approx linear}.In addition, the above results show that $\mbox{E}[\| s_{t} \|_{2}^{2}] < \infty$. The conclusion of the Lemma follows.

\subsection{Proof of Lemma \ref{Lemma2qmle}}

The score function $S_n^m$ for the perturbed model is a martingale sequence, with $\mbox{E}(S_n^m |{\cal F}_{t-1,m}^{\mathbold{Y}, \mathbold{\lambda}})=S_{n-1}^m$
at the true value $\mathbold{\theta}_0$ and ${\cal F}_{t-1,m}^{\mathbold{Y}, \mathbold{\lambda}}$ denotes the $\sigma$-field generated by $\{\mathbold{Y}_0^m,...,\mathbold{Y}_{t-1}^m, \mathbold{\epsilon}_0^{m},...,\mathbold{\epsilon}_{t-1}^{m}
\}$. In the previous section, we have already shown that it is square integrable.
An application of the strong law of large numbers for martingales (\citet{Chow1967}) gives almost sure convergence to $0$ of $S_n^m/n$ as $n \rightarrow \infty$.
To show asymptotic normality of the perturbed score function $S_n^m$ we apply the CLT for martingales; see \citet[Cor. 3.1]{HallandHeyde(1980)}.
Indeed, $(S_n^m)_{n \geq 1}$ is a zero mean, square integrable martingale sequence with $(s_t^m)_{t \geq \mathbb{N}}$ a martingale difference sequence.
To prove the conditional Lindeberg's condition
note that
\begin{eqnarray*}
\frac{1}{n} \sum \limits_{t=1}^n
\mbox{E} \Bigl[ \| s_t^m \|_{2}^2  I( \| s_t^m \|_{2} > \sqrt{n} \delta ) | {\cal F}_{t-1,m}^{\mathbold{Y}, \mathbold{\lambda}} \Bigr]  \rightarrow 0,
\end{eqnarray*}
since $E||s_t^m||_{2}^4 < \infty$. In addition,
\begin{eqnarray*}
\frac{1}{n} \sum_{t=1}^n
\mbox{Var} \Bigl[ s_t^m  | {\cal F}_{t-1,m}^{\mathbold{Y}, \mathbold{\lambda}}  \Bigr]
\xrightarrow{p}&  \mathbold{G}^{m}.
\end{eqnarray*}
This concludes the second result of the Lemma.

\noindent
The third result of the Lemma  follows from Lemma \ref{Lemma1qmle}  by using   \citet[Prop. 6.3.9.]{BrockwellandDavis(1991)}
Consider now the last result of the Lemma.
\begin{eqnarray*}
\frac{1}{\sqrt{n}}(S_n^m-S_n) &=& \frac{1}{\sqrt{n}} \sum\limits_{t=1}^n \left\{ s_t^m - s_t \right\} \nonumber \\
& = &
\frac{1}{\sqrt{n}}\sum_{t=1}^{n}
\Bigl[ \Bigl( \frac{\partial \mathbold{\lambda}_{t}^{m}}{\partial \mathbold{\theta}^{T}} \Bigr)^{T}
      - \Bigl(\frac{\partial \mathbold{\lambda}_{t}}{\partial \mathbold{\theta}^{T}} \Bigr)^{T} \Bigr]
\Bigl( \mathbold{D}_{t}^{m} \Bigr)^{-1}
\Bigl( \mathbold{Y}_{t}^{m}- \mathbold{\lambda}_{t}^{m}  \Bigr)   \nonumber \\
&+&
\frac{1}{\sqrt{n}} \sum_{t=1}^{n}
\Bigl( \frac{\partial \mathbold{\lambda}_{t}}{\partial \mathbold{\theta}^{T}} \Bigr)^{T}
\Bigl[ \Bigl( \mathbold{D}_{t}^{m} \Bigr)^{-1} -\Bigl( \mathbold{D}_{t} \Bigr)^{-1} \Bigr]
\Bigl( \mathbold{Y}_{t}^{m}- \mathbold{\lambda}_{t}^{m} \Bigr) \nonumber  \\
& + &
\frac{1}{\sqrt{n}} \sum_{t=1}^{n}
\Bigl( \frac{\partial \mathbold{\lambda}_{t}}{\partial \mathbold{\theta}^{T}} \Bigr)^{T}
\mathbold{D}_{t}^{-1}
\Bigl[ \Bigl( \mathbold{Y}_{t}^{m}- \mathbold{\lambda}_{t}^{m} \Bigr)-\Bigl( \mathbold{Y}_{t}- \mathbold{\lambda}_{t}\Bigr) \Bigr]   \nonumber \\
\end{eqnarray*}
For the first summand in the above representation, we obtain that
\begin{eqnarray*}
{P}\left( \left\| \sum_{t=1}^{n}
\Bigl[ \Bigl( \frac{\partial \mathbold{\lambda}_{t}^{m}}{\partial \mathbold{\theta}^{T}} \Bigr)^{T}
      - \Bigl(\frac{\partial \mathbold{\lambda}_{t}}{\partial \mathbold{\theta}^{T}} \Bigr)^{T} \Bigr]
\Bigl( \mathbold{D}_{t}^{m} \Bigr)^{-1}
\Bigl( \mathbold{Y}_{t}^{m}- \mathbold{\lambda}_{t}^{m}  \Bigr)   \right\|_{2}
>\delta \sqrt{n}\right)
& \leq &  {P}\left(
\gamma_{m} \left\| \sum_{t=1}^{n}
\Bigl( \mathbold{D}_{t}^{m} \Bigr)^{-1}
\Bigl( \mathbold{Y}_{t}^{m}- \mathbold{\lambda}_{t}^{m}  \Bigr)  \right\|_{2}  >\delta \sqrt{n}\right) \\
& \leq &  \frac{\gamma _{m}^{2}}{\delta ^{2}n}\sum_{t=1}^{n} \mbox{E} \left\|
\Bigl( \mathbold{D}_{t}^{m} \Bigr)^{-1}
\Bigl( \mathbold{Y}_{t}^{m}- \mathbold{\lambda}_{t}^{m}  \Bigr)  \right\|_{2}^{2}  \\
& \leq &  C\gamma _{m}^{2}\rightarrow 0,
\end{eqnarray*}
as $m \rightarrow \infty$, for some $\gamma_{m}$.
The other two terms are treated similarly given that
$E \Bigl\|\Bigl| \partial  \mathbold{\lambda}_t / \partial \mathbold{\theta}^T \Bigr\| \Bigr|_{2}^2< \infty$ which has already been proved in the previous arguments.

\subsection{Proof of Lemma \ref{Lemma3qmple}}

The proof of this Lemma is based on identical arguments given in the proof of  \citet[Lemma 3.3]{Fokianosetal(2009)}
and therefore it is omitted. Positive definiteness of the matrix $\mathbold{H}$ follows directly from \eqref{Hessian limit} since $\lambda_{i,t}  > d_{i}$ where
$d_{i}$ is the $i$'th component of vector $\mathbold{d}$.

\subsection{Remarks about third derivative calculations for model \eqref{linear model mult}}

The third order derivative of the log-likelihood terms
$$
l_{t}\left(  \boldsymbol{\theta}\right) = \sum_{i=1}^{p} l_{i,t}(\mathbold{\theta}),
$$
where $l_{i,t}(\mathbold{\theta})= y_{i;t} \log \lambda_{i,t}(\mathbold{\theta})- \lambda_{i;t}(\mathbold{\theta})$, $i=1,2,\ldots,p$
are given by   ${\partial^{3}l_{t}\left(  \boldsymbol{\theta}\right)}/{\partial\boldsymbol{\theta}_{l}\partial\boldsymbol{\theta}_{j}\partial\boldsymbol{\theta}_{k}}$
with
\begin{align*}
\displaystyle
\frac{\partial^{3}l_{i,t}\left(  \boldsymbol{\theta}\right)  }{\partial\boldsymbol{\theta}_{l}%
\partial\boldsymbol{\theta}_{j}\partial\boldsymbol{\theta}_{k}}  &  =-\left(  \frac{Y_{i,t}}{\lambda
_{i, t}^{2}\left(  \boldsymbol{\theta}\right)  }\right)  \left(  \frac{\partial^{2}%
\lambda_{i,t}\left(  \boldsymbol{\theta}\right)  }{\partial\boldsymbol{\theta}_{l}\partial\boldsymbol{\theta}_{j}%
}\frac{\partial\lambda_{i, t}\left(  \boldsymbol{\theta}\right)  }{\partial\boldsymbol{\theta}_{k}}%
+\frac{\partial^{2}\lambda_{i,t}\left(  \boldsymbol{\theta}\right)  }{\partial\boldsymbol{\theta}
_{l}\partial\boldsymbol{\theta}_{k}}\frac{\partial\lambda_{i,t}\left(  \boldsymbol{\theta}\right)
}{\partial\boldsymbol{\theta}_{j}}+\frac{\partial^{2}\lambda_{i,t}\left(  \boldsymbol{\theta}\right)
}{\partial\boldsymbol{\theta}_{j}\partial\boldsymbol{\theta}_{k}}\frac{\partial\lambda_{i,t}\left(
\boldsymbol{\theta}\right)  }{\partial\boldsymbol{\theta}_{l}}\right) \\
&  +2\left(  \frac{Y_{i,t}}{\lambda_{i,t}^{3}\left(  \boldsymbol{\theta}\right)  }\right)
\left(  \frac{\partial\lambda_{i,t}\left(  \boldsymbol{\theta}\right)  }{\partial\boldsymbol{\theta}_{k}%
}\frac{\partial\lambda_{i,t}\left(  \boldsymbol{\theta}\right)  }{\partial\boldsymbol{\theta}_{j}}%
\frac{\partial\lambda_{i,t}\left(  \boldsymbol{\theta}\right)  }{\partial\boldsymbol{\theta}_{l}}\right)
+\left(  \frac{Y_{i,t}}{\lambda_{i,t}\left(  \boldsymbol{\theta}\right)  }-1\right)
\frac{\partial^{3}\lambda_{i,t}\left(  \boldsymbol{\theta}\right)  }{\partial\boldsymbol{\theta}
_{l}\partial\boldsymbol{\theta}_{j}\partial\boldsymbol{\theta}_{k}}.
\end{align*}
Then all these terms can be bound suitably along the arguments of \citet[Lemma 3.4]{Fokianosetal(2009)}.

\subsection{On local Gaussian correlation}
\label{Sec:LGC}

 Let $X=(X_1,X_2)$
be a two-dimensional random variable with density $f(x)=f(x_1,x_2)$. In this section we describe how $f$ can be approximated locally in a
neighbourhood of each point $x=(x_1,x_2)$ by a Gaussian bivariate density of the
\begin{equation}\label{psi1}
\psi(v,\mu(x),\Sigma(x))=\dfrac{1}{2\pi |\Sigma(x)|^{1/2}}{\exp}\left[-\frac{1}{2}(v-\mu(x))^T\Sigma^{-1}(x)(v-\mu(x))\right],
\end{equation}
where $v=(v_1,v_2)^T$, $\mu(x)=(\mu_1(x),\mu_2(x))^T$ is
the local mean vector and $\Sigma(x)=(\sigma_{ij}(x))$ is the local covariance
matrix. With $\sigma_{i}^2(x)=\sigma_{ii}(x)$, we define the local Gaussian correlation
at the point $x$ by $\rho(x)=\frac{\sigma_{12}(x)}{\sigma_1(x)\sigma_2(x)}$. Then \eqref{psi1} becomes
\begin{multline}\label{gaussiandistr2}
\psi(v,\mu_1(x),\mu_2(x),\sigma_{1}^2(x),\sigma_{2}^2(x),\rho(x))=\\
\dfrac{1}{2\pi\sigma_1(x)\sigma_2(x)\sqrt{1-\rho^2(x)}}\exp\Biggr\lbrace
-\dfrac{1}{2(1-\rho^2(x))} \times
\Biggr[\biggr(\dfrac{v_1-\mu_1(x)}{\sigma_1(x)}\biggr)^2\\
-2\rho(x)\biggr(\dfrac{v_1-\mu_1(x)}{\sigma_1(x)}\biggr)\biggr(\dfrac{v_2-\mu_2(x)}{\sigma_2(x)}\biggr)+\biggr(\dfrac{v_2-\mu_2(x)}{\sigma_2(x)}\biggr)^2\Biggr]\Biggr\rbrace.
\end{multline}
\noindent
First note that  \eqref{gaussiandistr2} is not
well-defined unless some extra conditions are imposed.
We need
to construct a Gaussian approximation that approximates $f(x)$ in a {\it
neighborhood} of $x$ and such that \eqref{gaussiandistr2} holds at $x$.
In \citet{TjostheimandHufthammer(2013)} it was shown  that for a given neighbourhood
characterized by a bandwidth parameter $b$ the local population parameters
$\gamma(x) = (\mu(x),\Sigma(x))$ or $\gamma(x) = (\mu_1(x),\mu_2(x),\sigma_1^2(x),\sigma_2^2(x), \rho(x))$ can be defined by
minimizing a likelihood related penalty function $q$ given by
\begin{equation}\label{eq:q}
q=\int K_b(v-x)
\left[\psi(v,\gamma(x))-\log\psi(v,\gamma(x))f(v)\right]dv,
\end{equation}
where  $K_b(v-x) =
b^{-1}K(b^{-1}(v-x))$ with $K$ being a kernel function. We define the population value
$\gamma(x)=\gamma_b(x)$ as the minimizers of this penalty function. It
then satisfies the set of equations
\begin{equation}
\int K_b(v-x) \frac{\partial}{\partial
\gamma_j}\log\psi(v,\gamma(x))\left[f(v)-\psi(v,\gamma(x))\right]dv = 0,
\;\;j=1,\ldots,5. \label{eq:intzero}
\end{equation}
\noindent
\citet{TjostheimandHufthammer(2013)} show  that if  a unique
population vector $\gamma_b(x)$ exists then, under weak regularity conditions,  one can
let $b \to 0$ to obtain a local population vector $\gamma(x)$ defined at a point $x$. The population vectors
$\gamma_b(x)$ and $\gamma(x)$ can both be consistently estimated
by using a local log-likelihood function defined by
\begin{equation}
  L\bigl(X_1,\dots,X_n,\gamma_b(x)\bigr)\\
=n^{-1}\sum_i K_b(X_i-x)\log\psi(X_i,\gamma_b(x)) -\int
K_b(v-x)\psi(v,\gamma_b(x))  dv,\label{eq:locallik}
\end{equation}
for given observations $X_1,\ldots,X_n$, see \citet{HjortandJones(1996)}.
Numerical maximization of the local likelihood (\ref{eq:locallik}) leads to
local likelihood estimates $\gamma_{n,b}(x)$, including estimates
$\rho_{n,b}(x)$ of the local Gaussian correlation. It is shown in
\citet{TjostheimandHufthammer(2013)} that, under relatively weak regularity conditions,
$\gamma_{n,b}(x) \to \gamma_b(x)$ for $b$ fixed, and $\gamma_{n,b}(x) \to
\gamma(x)$ almost surely for $b=b_n$ tending to zero. In addition asymptotic normality is demonstrated in that paper. Further,
equation (\ref{eq:locallik}) is consistent with (\ref{eq:q}).

\subsection{Copula Estimation}
\label{sec:copula estimation}

An additional estimation problem  is the estimation of dependence among the observed time series. This is equivalent to the
problem of obtaining an  estimate for the copula. In this appendix, we give a heuristic method for achieving this by specifying a parametric copula form.
A thorough study of this problem  including asymptotic properties of estimates will be reported elsewhere.

More precisely, we are suggesting a parametric bootstrap based algorithm for identification of  the copula structure, which is assumed to be of the form  $C_{\phi}(\cdot)$, where
$\phi$ is an unknown copula parameter. We restrict the methodology  to one-parameter copulas.
For simplicity, we only outline  the algorithm for the bivariate case but
a multivariate extension is possible under some further  assumptions. The proposed  algorithm  employs
the theory of local Gaussian correlation (LGC),  as presented by
\citet{TjostheimandHufthammer(2013)}. LGC is a local correlation measure that can give a precise description of non-linear dependence between variables; see \ref{Sec:LGC} for more details.
In \citet{Berentsenetal(2014)} it is shown that the LGC is  able to capture  characteristics of the dependence structure of different copula models in the continuous case in
a very satisfactory manner. These encouraging results show that its use might be potentially useful for identifying a continuous copula structure approximately in the present case of discrete
variable $(Y_{1,t}, Y_{2,t})$ with distribution being approximated by a continuous distribution.
The parametric bootstrap procedure for a bivariate count time series is given as  follows:

\begin{itemize}
 \item[1.] Given  the observations $Y_{1,t}$ and $Y_{2,t}$, $t=1,...,n$ estimate $\hat{\mathbold{\theta}}$ and $\hat{\lambda}_{1;t}$ and $\hat{\lambda}_{2;t}$  for $t=1,2,\ldots,n$.
 \item[2.] For a given copula structure and for a given value of the copula parameter, generate a sample of bivariate Poisson variables $Y_{1,t}^*$ and $Y_{2,t}^*$, $t=1,...,n$  by the data generating mechanism given in section 2 using the estimates from step 1.
 \item[3.] Compute the local Gaussian correlation $\rho_{n,b}(\cdot)$ between $Y_{1,t}$ and $Y_{2,t}$, and the local Gaussian correlation
 $\rho_{n,b}^*(\cdot)$ between $Y_{1,t}^*$ and $Y_{2,t}^*$ on a pre-defined grid $(u_j, v_j)$, $j=1,...,m$.
 \item[4.] Compute the distance measure $D_m = \frac{1}{m} \sum_{j=1}^m [\rho_{n,b}(u_j, v_j) - \rho_{n,b}^*(u_j, v_j)]^2$.
 \item[5.] Repeat steps 2 to 4 for different copula structures and over a grid of values for the copula parameter. The estimate of the copula structure
 and corresponding copula parameter, $\hat{\phi}$, is the one that minimizes $D_m$.
 \end{itemize}
 If we are interested  in the standard error of the estimate for the copula parameter, then we  have to include the additional step:
 \begin{itemize}
 \item[6.] Repeat steps 2 to 5 $B$ times to obtain $\hat{\phi}_{1}, \ldots, \hat{\phi}_{B}$ by selecting the copula structure and parameter value
 that minimize $D_m$ for each realization. The copula
parameter $\phi$ is estimated by the average of these realizations (by only considering the realizations for the copula structure that is selected most of the times).
In addition, its standard error is obtained by simply considering the standard error of those realizations.
\end{itemize}

This algorithm will be employed and studied in Section \ref{sec:simresult:copula} of the Appendix  where we give several examples.

\subsection{Some simulation results on copula estimation}
\label{sec:simresult:copula}

To carry out the algorithm given in Section \ref{sec:copula estimation}, we generate  100 realizations from the  bivariate process  defined by the
linear model \eqref{linear model mult} with parameter values given by \eqref{first linear specification}
and then execute  step 1 of the algorithm, with sample sizes of 500 and 1000. Then  we do
steps 2 to 5 to  select the estimator with the minimum value of the distance function $D_m$ (see step 5), for each of the realizations.

Moreover,  we perform two separate simulations by considering  different copula structures.
In the first case we use a Gaussian copula with parameter $\phi = 0.5$ and
in the second case we use a Clayton copula with parameter $\phi = 4$. Implementation of  step 5 of the algorithm is  done as follows.
For the Clayton copula, we generate  a grid of
the parameter value $\phi$    from  0.5 to 8. For the  Gaussian copula case the  grid is chosen by varying $\phi$  from -1 to 1.
The grid for calculating the LGC is chosen to be diagonal, i.e.
$u_j = v_j$ starting from 1 and up to the maximum value in the generated Poisson process in step 1.
The bandwidths are calculated as the standard deviation
of the same process multiplied with 1.1. We thus prefer  to oversmooth slightly, and this is a reasonable bandwidth choice which has been
advocated by  \citet{Stoveetal(2014)}. Although the  LGC is theoretically only defined for continuous random variables,
it is still interesting to investigate  whether  it gives reasonable results  for  discrete data,  at least empirically.

The simulation results are given in Table \ref{tab:6}. The number in squared
brackets indicate how many of the realizations have chosen  the correct copula,
and  the estimates of the copula parameter are then found by averaging out the results from those realizations.
Similarly, the standard errors reported correspond to the sampling standard errors of the estimates obtained by the same realizations.
The results are quite satisfactory. For the Clayton copula, the algorithm  chooses
the correct copula structure in 88 and 94 times (out of 100) for sample sizes 500 and 1000, respectively.
The estimated copula parameter is relatively close to its true value; in particular when $n=1000$.
For the Gaussian copula, the estimated copula parameter is almost identical to its true value,
even though the procedure is slightly less accurate than the procedure for the Clayton copula. Indeed, the algorithm chooses the correct copula
structure   69 and 70 times (out of 100), for  sample size of 500 and 1000, respectively.

\begin{table}[htbp] \small
 \centering
 \begin{tabular}{c|cc}\hline \hline
Sample size & Clayton copula with $\phi=4$  & Gaussian copula with $\phi=0.5 $  \\ \hline
\multirow{2}{*}{500} &  5.17 [88] &  0.48 [69]  \\
 &   (1.59) & (0.15)       \\ \hline
\multirow{2}{*}{1000} &  4.51 [94] & 0.49 [70]   \\
 &  (1.50)& (0.11) \\ \hline
\end{tabular}
\caption{Simulation results for the parametric copula estimation algorithm  given  in Section \ref{sec:copula estimation}. Data are generated
by the linear model \eqref{linear model mult} with parameter values  given by \eqref{first linear specification}. Results are based on 100 runs. }
\label{tab:6}
\end{table}

\subsection{Copula estimation for real data}
\label{sec:copestimrealdata}

First, we note that the standard correlation coefficient
between the two series is estimated to be 0.358 (see also Figure \ref{fig:auto}).
We also estimate the LGC between the series by using \eqref{eq:locallik} in a diagonal grid; see  Figure \ref{fig:lgc}.
The LGC is between 0.2 and 0.35, and it attains its  highest  value around a count of 6.
Clearly, the dependence pattern is highly non-linear, and possibly not easily captured.
However, we  proceed using the proposed parametric bootstrap routine as was explained earlier. In particular, we only let the algorithm
choose between the Gaussian and Clayton copula. In this case we also utilize step 6 of  the algorithm, in order to estimate the standard error of the
copula parameter. Note that we consider   $B = 1000$. The estimation
procedure selects the Clayton copula in 68 (respectively 51) cases out of the 100 for the linear (respectively, log-linear) model.
For the linear model case, the estimated copula parameter is 2.65 with standard error of 0.95. For the log-linear model
the estimated copula parameter is  2.01 with standard error equal to 0.81.
The parameter value seems rather large  when comparing to the plot of  Figure \ref{fig:lgc}. Furthermore, the standard error
of the copula parameter is relatively large, indicating that the estimation problem in this particular case is challenging and that the Clayton copula
may not be optimal.

\begin{figure}[htb]
 \centering
 \includegraphics[scale=0.40]{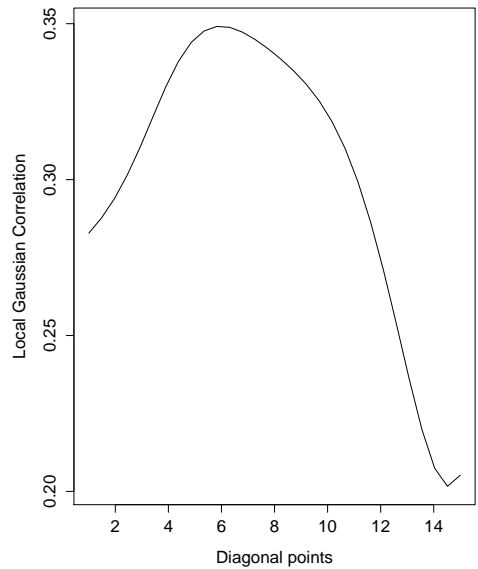}
 \caption{The estimated local Gaussian correlation between the IBM and Coca-Cola transaction data on a diagonal grid.}
\label{fig:lgc}
\end{figure}


\end{document}